\documentclass[12pt]{iopart}
\usepackage[dvips]{graphicx}
\usepackage{longtable}
\usepackage{setstack}
\usepackage{iopams}
\usepackage{color}
\usepackage{amssymb}
\usepackage{cite}                      % "collapse" citations [1,2,3,5] -> [1-3,5]
\newcommand{\Fig}[1]{figure~\ref{#1}}  % THIS IS NEEDED !
\begin{document}
\topical{Charge transport through single molecules, quantum dots, and quantum wires}

\author{S. Andergassen$^1$, V. Meden$^1$, H. Schoeller$^1$, J. Splettstoesser$^1$ and M.R. Wegewijs$^{1,2}$}

\address{$^1$ Institut f\"ur Theoretische Physik A, RWTH Aachen, 
  52056 Aachen, Germany, and JARA-Fundamentals of Future Information Technology}
\address{$^2$ Institut f\"ur Festk\"orperforschung - Theorie 3,
  Forschungszentrum J\"ulich, 52425 J\"ulich, Germany}
\ead{schoeller@physik.rwth-aachen.de}
\begin{abstract}
We review recent progresses in the theoretical description of correlation
and quantum fluctuation phenomena in charge transport through
single molecules, quantum dots, and quantum wires.
A variety of physical phenomena is addressed, relating to co-tunneling,
pair-tunneling, adiabatic quantum pumping, charge and
spin fluctuations, and inhomogeneous Luttinger liquids.
We review theoretical many-body methods to treat correlation effects,
quantum fluctuations, nonequilibrium physics, and the time evolution
into the stationary state of complex nanoelectronic systems.

\end{abstract}

% Uncomment to check for approximate two column behavior:
%\twocolumn

\maketitle

\section{Introduction}
\label{sec:intro}
In this review we consider different aspects of charge transport through nanoscale 
devices attached to electronic reservoirs,
focusing on theoretical approaches dealing with interaction and quantum fluctuation effects.
Transport experiments have shown convincingly that many of these systems - from carbon 
nanotubes down to single nanometer sized molecules -  behave as \emph{quantum dots}.
A quantum dot is a confined system of electrons, which is so small that the discreteness 
of energy spectrum with a level spacing $\Delta\epsilon$ becomes important;
it is therefore often referred to as an artificial atom. 
Due to the smallness of the quantum dot Coulomb interaction between the electrons also
plays an important role: the charging energy $U$ which can be of same
order as the level spacing.
If the quantum dot is attached to reservoirs by tunneling contacts, electrons can 
leave and enter the dot and the single-particle levels are therefore broadened due to the finite 
lifetime of the electrons. For sufficiently weak tunnel coupling $\Gamma$ and small
temperature $T$, the level spacing $\Delta\epsilon$ and the charging energy $U$ can be 
resolved at standard cryogenic temperatures. This opens up the possibility of 
performing transport spectroscopy by measuring the differential conductance as function of 
gate voltage $V_{\mathrm{g}}$ and bias voltage $V$.
The transport characteristics allow to count the discrete charge states of such systems, 
and even more complex degrees of freedom of the quantum dot are revealed.  As an example, 
the spin states can be identified with the help of the Zeeman effect, and also vibrational 
motion of the quantum dot system can be detected.

The transport behavior, however, depends strongly on the ratios of all the energy scales,
and on the way transport is induced, e.g., by static or time-dependent applied voltages, 
or by measuring linear or non-linear response in the transport voltage.
In particular, for molecular scale quantum dots the excitation spectra may be quite complex 
and various splittings appear. This complicates matters further but also leads to a variety 
of transport effects.
Besides stationary properties of quantum dots, the time evolution out of an initially
prepared nonequilibrium state is another important issue of recent interest, providing the
new energy scale $\hbar/t$. The time evolution itself can be used as a tool to reveal 
various many-body aspects. It is also of practical importance
due to the progress in the field of solid-state quantum information processing, initiated by the 
suggestion to use spin states of quantum dots as basic elements of qubits \cite{Loss98}.
In addition to the  ``bare'' energy scales mentioned so far, due to correlation effects 
\emph{new effective energy scales} can be generated, which describe the physics of quantum fluctuations 
induced by the reservoir-dot coupling. These new energy scales are complex functions of the above 
``bare'' energies and we denote them in the following by $T_{\mathrm{c}}$, where c stands for ``cutoff''.
Examples are the Kondo temperature describing the strong coupling scale for spin 
fluctuations (usually denoted by $T_{\mathrm{K}}$) and renormalised tunneling
couplings generated by charge fluctuations. 
These scales will dominate the measured transport in the strong coupling regime 
where $T_{\mathrm{c}}$ is the largest energy scale. However, also in the weak coupling
regime quantum fluctuations set on and are of particular interest close to resonances, where
they can be analysed by a perturbative treatment in the renormalised couplings.
To understand the new energy scales physically and calculate the renormalised couplings, 
\emph{renormalisation group} (RG) ideas are required. RG transport theories reveal that interactions 
are of key importance for the generation of new effective energy scales
by quantum fluctuations. Without interactions, quantum fluctuations are well understood,
they just lead to an unrenormalised broadening of the energy conservation by the 
tunneling coupling. Although new effects from correlations 
are most pronounced in the regime of strong correlations $\Gamma\ll U$, 
interesting quantum fluctuation effects already start to become visible in the regime of moderate 
correlations $\Gamma\sim U$, where perturbative and mean-field theories are often not yet applicable.
Another interesting regime is the case of a continuous level spectrum,
i.e., where the quantum dot is replaced by a one-dimensional quantum wire. In this case, 
quantum fluctuations in the wire itself lead to renormalisation effects and 
Luttinger liquid physics, with typical power-law suppression of the bulk spectral density. 
Similar to the case of quantum dots, the {\it simultaneous} presence of correlations and transport
through inhomogeneities leads to very rich phenomena. A prominent example is the suppression of 
the spectral density at wire boundaries, which leads to vanishing conductance at zero temperature 
and bias voltage if a quantum wire is weakly coupled to two leads. This has to be contrasted 
to the Kondo effect in quantum dots, where conductance becomes maximal when a single spin is placed 
in an ``insulating'' quantum dot.

In this review we will illustrate these key issues and concentrate on correlation
effects in quantum dots and quantum wires. We will describe transport theories 
specifically designed to treat the case of moderate to strong Coulomb interactions
and quantum fluctuation effects. 
With these methods it is possible to account for the complex interplay of various 
combinations of quantum mechanical effects (level quantization, interference, 
spin and angular momentum), complex excitations (e.g., due to spin-orbit interaction 
and/or mechanical vibrations), Luttinger liquid physics, 
non-equilibrium (due to static as well as time-dependent applied voltages), and the 
time evolution into the stationary state. An important aspect is that these methods 
apply to general models incorporating experimentally relevant microscopic details,
and offer prospects for numerical implementation. Despite these common themes, the 
different sections of this review are meant to be almost self-contained and each section 
closes with a short outlook. We consistently use natural units $\hbar=k_{\mathrm{B}}=e=1$.
The paper is organized as follows.

In section \ref{sec:molecules}, we will introduce a general approach to transport through quantum 
dots of arbitrary complexity,
aiming at the description of three terminal single-molecule junctions.
The tunneling coupling is treated perturbatively up to second order in the coupling $\Gamma$,
accounting for charge fluctuation effects, while not capturing spin fluctuations which start in higher order.
Besides well-known level renormalisation and cotunneling effects,
electron pair tunneling is shown to give rise to resonances, even for the most basic quantum dot model.
It is shown how the complexity of, e.g., vibrational and magnetic excitations of molecular quantum dots
needs to be accounted for in transport spectroscopy calculations.

Section \ref{sec:external_fields} uses the same perturbation theory but extends it to transport 
through systems which are subject to time-dependent fields. Using the example of adiabatic charge 
pumping through a single-level quantum dot, it explains why time-dependent external fields in the 
adiabatic regime are of particular interest. It is shown how time-dependent fields 
can be used to reveal level renormalisations induced by charge fluctuations, an effect purely due to 
the Coulomb interaction. This offers the interesting perspective to use time-dependent fields as
fingerprints for the combined influence of quantum fluctuations and correlations.

The regime of strong quantum fluctuations is addressed
in section \ref{sec:fluct_linear} in the linear response regime. It is described
how a recently developed functional RG (fRG) method is a useful tool to treat 
generic quantum dot models. Within this method a systematic expansion in 
a renormalised Coulomb interaction is performed. However, the tunneling coupling is treated 
non-perturbatively, so that the strong coupling regime can be addressed for
spin as well as for charge fluctuations, at least for moderate Coulomb interactions.
Two prominent examples of strong quantum fluctuations are reviewed: The Kondo 
effect in quantum dots and the explanation of transmission phase lapses by
strong broadening and renormalisation effects in multi-level quantum dots.
The latter involves a significant progress in the insight for 
an important long-outstanding problem posed by experiments.

Section \ref{sec:fluct_nonlinear} describes quantum fluctuations 
in nonlinear response, i.e. at finite bias voltage $V\gg T$, together with the time evolution 
into the stationary state. The basic physics of weak spin and strong charge fluctuations
is illustrated  on recent results for two elementary models:
the anisotropic Kondo model at finite magnetic field and the
interacting resonant level model. A real-time renormalisation group (RTRG) method is 
introduced, where, complementary to the fRG scheme presented in section \ref{sec:fluct_linear},
an expansion in a renormalised tunneling coupling is performed whereas the Coulomb interaction
is treated non-perturbatively. This allows the description of strong correlation effects
$\Gamma\ll U$ for moderate tunneling.

Finally, in section \ref{sec:wires} we describe transport properties of quantum wires.
The status of Luttinger liquid physics is reviewed, in particular in connection with the presence of 
inhomogeneities, backscattering, interference effects, and nonequilibrium. As in
section \ref{sec:fluct_linear}, the fRG scheme is shown to be a unique method, capable of describing
the whole crossover from few- to multi-level dots up to the limit of continuous spectra in quantum wires.
Its ability to simultaneously incorporate microscopic details is of particular importance for the 
description of experiments.

%%%%%%%%%%%%%%%%%%%%%%%%%%%%%%%%%%%%%%%%%%%%%%%%%%%%%%%%%%%%%%%%%%%%%%%
\section{Transport through single molecule quantum dots}
\label{sec:molecules}
Molecular quantum dots present the ultimate miniaturization of quantum dot devices,
which have been realized in semi-conductor heterostructures and wires,
metallic nanoparticles, and carbon-nanotube molecular wires.
Various methods have been developed to contact nanometer size single molecules.
Transport measurements in three terminal junctions have most clearly demonstrated that single molecules
also are quantum dots, although of a particularly complex nature, 
see for a review~\cite{Natelson06,vanderZant06,Osorio08rev}.
% What is the challenge ?:
In view of this a central theoretical challenge is to describe the transport through a quantum dot of arbitrary complexity, covering the entire, broad class of systems mentioned above.
We now first outline this general framework, which is relevant also for 
sections~\ref{sec:external_fields} and \ref{sec:fluct_nonlinear}.
For details see~\cite{Schoeller09a} and the references therein.

Starting from the limit of completely opaque tunnel junction, we can specify the exact many-body level spectrum and states $|s\rangle$ of the quantum dot in this case,
\begin{eqnarray}
  \label{eq:HD}
  H_\mathrm{D} & = & \sum_{s} E_s |s\rangle \langle s|
  .
\end{eqnarray}
Often, only a few of these states are actually accessible in an experiment.
Subsequently, we incorporate the non-zero transparency of the junction in the coupling
\begin{eqnarray}
  \label{eq:HT}
  H_T & = & \sum_{s,s'} \sum_{\alpha k\sigma} T^{s s'}_{\alpha k\sigma} |s\rangle \langle s'| c_{\alpha k \sigma} + \mathrm{h.c.}
\end{eqnarray}
through the amplitudes $T^{s s'}_{\alpha k\sigma}$ for an electron with spin $\sigma$ to tunnel on / off the dot
from / to one of the electronic reservoirs labeled by $\alpha=\mathrm{L,R}$.
Generally, in \Eref{eq:HT} all states $s$ and $s'$ are coupled which differ by the addition of one electron.
The reservoir $\alpha$ by itself is described by
\begin{eqnarray}
  \label{eq:Halpha}
  H_\alpha & = & \sum_{k\sigma} (\epsilon_{k\sigma}+\mu_\alpha)n_{\alpha k \sigma}
  ,
\end{eqnarray}
where $n_{\alpha k \sigma}$ denotes the electron number operator and $k$ labels the orbital states.
Each electrode is restricted to contain effectively non-interacting fermion quasi-particles with density of states $\rho_\alpha$.
We make the further statistical assumption that the electro-chemical potential $\mu_\alpha$ and temperature $T$ are well defined for each reservoir $\alpha$ separately.
The central quantity is the reduced density operator $p$ of the dot, obtained by partial averaging of the full density operator over the reservoirs.
From $p$ any non-equilibrium expectation value of a local observable can be calculated,
and also (see below) the transport current.
Starting from the above general model, one can derive the kinetic equation from which the time-evolution of the dot density operator can be calculated:
\begin{eqnarray}
  \label{eq:kineq}
  \frac{d}{dt}p(t) & = & -\rmi L_{\mathrm{D}}~p(t) -i \int_{t_0}^{t} dt'\,\Sigma(t,t') ~ p(t')
  .
\end{eqnarray}
Here the Liouvillian superoperator $L_{\mathrm{D}}$ acts on a density operator by commuting it  with $H_{\mathrm{D}}$, i.e., $L_{\mathrm{D}} p(t) = [H_{\mathrm{D}},p(t)]$.
This linear transformation generates the ``free'' time-evolution of the dot density matrix, i.e., in the case where it is not coupled to the electrodes.
The challenge lies in the calculation of the transport kernel $\Sigma(t,t')$ which incorporates all effects due to the coupling to the electrodes
which is adiabatically switched on at time $t_0$ ($t_0 < t$).
The superoperator $\Sigma(t,t')$ is a non-trivial functional of three ``objects'':
(i) the dot Liouvillian $L_{\mathrm{D}}$ (energy differences of initial, final and virtual intermediate states of processes),
(ii) the dot part of the tunnel operator (tunnel vertices) and
(iii) the product of the Fermi distribution functions $f(( \epsilon_{k\sigma}  -\mu_\alpha)/T)$ and density of states $\rho_\alpha$ of all electrodes $\alpha$.
Importantly, compact diagrammatic rules for the calculation of this kernel in terms of these three objects are known exactly~\cite{Schoeller09a,Leijnse08a}.
In this section and section~\ref{sec:external_fields} we review the basic physics which emerges from the calculation of the kernel by perturbation theory in the tunnel operator $H_T^2 \sim \Gamma$ to leading and next to leading order.
This is applicable to the ``high-temperature'' regime, where higher order effects are present but still weak.
In section ~\ref{sec:fluct_linear} and ~\ref{sec:fluct_nonlinear} the low-temperature physics due to strong quantum fluctuations is illustrated.
In particular, in section~\ref{sec:fluct_nonlinear} the kernel is calculated non-perturbatively in a renormalisation-group framework.
The technical details of these two approaches can be found in~\cite{Leijnse08a} and~\cite{Schoeller09a} respectively.

We now first review all possible basic types of transport resonances which can arise due to tunneling processes of the first and second order in $\Gamma$.
For this purpose we consider a single energy level with Coulomb interaction,  the so-called Anderson model,
where we emphasize the need to consider a \emph{finite} charging energy $U$~\cite{Leijnse09a} such that double occupancy is possible
 (not only as a virtual intermediate state).
This will set the stage for the subsequent discussion of examples of transport involving more complex spectra in section~\ref{sec:complex}
and the time-dependent transport considered in section~\ref{sec:external_fields}.

\subsection{Transport spectroscopy: sequential, co- and pair-tunneling}
\label{sec:spectro}
%%%%%%%%%%%%%%%%%%%%%%%%%%%%%%%%%%%%%%%%%%%%%%%%%%%
\begin{figure}
  \begin{center}
    \includegraphics[width=0.50\linewidth]{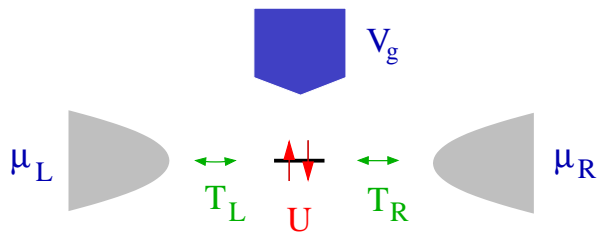}
  \end{center}
  \caption{
    (Color online)
    Sketch of a single-level quantum dot attached to two reservoirs with $\mu_{\mathrm{L/R}}=\pm V/2$.}
  \label{fig:skizze}
\end{figure}
%%%%%%%%%%%%%%%%%%%%%%%%%%%%%%%%%%%%%%%%%%%%%%%%%%%
Most types of transport resonances observed in quantum dot experiments can be illustrated
by the Anderson model sketched in figure~\ref{fig:skizze} where a single orbital level is coupled to electrodes $\alpha=\mathrm{L,R}$ with tunnel amplitudes $T_\alpha$.
The strength of the tunnel coupling is characterized by the spectral function, $\Gamma_\alpha = 2\pi\rho_\alpha| T_\alpha |^2$, whose energy dependence is neglected.
Denoting by $N$ the electron number on the dot, the eigenstates $|s\rangle$ of the dot (when decoupled from the electrodes) are easily classified.
For the empty dot ($N=0$) there is one state $s=0$ with energy $E_0=0$.
For the singly occupied dot ($N=1$) there are two states $s=\sigma=\uparrow,\downarrow$ with energy $E_\sigma =\varepsilon_\sigma$.
The $N=1$ states are split by a magnetic field such that $s=\downarrow$ is the ground state 
and $s=\uparrow$ the excited state.
Finally, for the doubly occupied dot ($N=2$) the energy  $E_\mathrm{d}=\sum_\sigma \varepsilon_\sigma+U$ of the state $s=\mathrm{d}$
 includes the charging energy.
One can show that due to conservation of the electron number and the spin-projection along the magnetic field only the occupation probabilities enter into the description of the non-equilibrium state.
We collect these into a vector
$\bi{p}=  (p_0,p_\uparrow,p_\downarrow,p_{\mathrm{d}})^{\mathrm{T}}$.
Its time-evolution is fully described by a kinetic equation with a matrix kernel $\boldsymbol{\Sigma}\left(t,t'\right)$,
\begin{eqnarray}
\label{eq_general_master}
	\frac{d}{dt}\bi{p}\left(t\right) & = & -\rmi~\int_{-\infty}^{t}dt'\boldsymbol{\Sigma} \left(t,t'\right)\bi{p}\left(t'\right)\
        ,
\end{eqnarray}
where we have set $t_0 =-\infty$ (c.f.~\eref{eq:kineq}).
For time-independent parameters considered here, we have $\Sigma(t,t')=\Sigma(t-t')$ and
the occupancies will reach a constant value in the stationary long-time limit:
$\bi{p}\left(t\right) \rightarrow \bi{p} =$~constant for $t-t_0 \rightarrow \infty$.
With the subsidiary condition of probability conservation $\sum_s p_{s}=1$, this state is determined uniquely by
\begin{eqnarray}
  \label{eq:stationary_master}
  0 & = & -\rmi~\boldsymbol{\Sigma}~\bi{p}
  .
\end{eqnarray}
As expected, here the time integral over the kernel is needed, i.e., the zero-frequency Laplace transform
$\lim_{z\rightarrow \rmi 0} \int_{0}^{\infty }dt e^{\rmi z t}\boldsymbol{\Sigma} \left(t,0\right) = \boldsymbol{\Sigma}$.
This kernel has been calculated in closed form up to order $\Gamma^2$ for an arbitrary dot model Hamiltonian in~\cite{Leijnse08a},
allowing arbitrarily complex transport spectra to be calculated.
From a similar equation for the measurable tunnel current from electrode $\alpha$
\begin{eqnarray}
\label{eq_current}
	I_{\alpha} =
        -  \rmi  \int_{-\infty}^{t}dt' \,\mathrm{Tr}_{\mathrm{D}}
\boldsymbol{\Sigma}^{\alpha}\left(t,t'\right)\bi{p}\left(t'\right)
        \rightarrow
         - \rmi ~ \mathrm{Tr}_{\mathrm{D}} ~ \boldsymbol{\Sigma}^{\alpha} ~ \bi{p},
\end{eqnarray}
the stationary current can be found.
This requires an additional zero-frequency kernel $\boldsymbol{\Sigma}^{\mathrm{\alpha}}$ which takes account of both the amount and the direction of the electrons transferred from electrode $\alpha$.
Here $\mathrm{Tr}_{\mathrm{D}}$ denotes the trace over the dot degrees of freedom.
%the vector $\bi{n}^{\mathrm{T}}=(1,...,1)$ merely serves to sum up all the vector 
%elements of  $\boldsymbol{\Sigma}^{\mathrm{\alpha}}\bi{p}$.
%%%%%%%%%%%%%%%%%%%%%%%%%%%%%%%%%%%%%%%%%%%%%%%%%%%
\begin{figure}
  \begin{center}
    \includegraphics[width=0.80\linewidth]{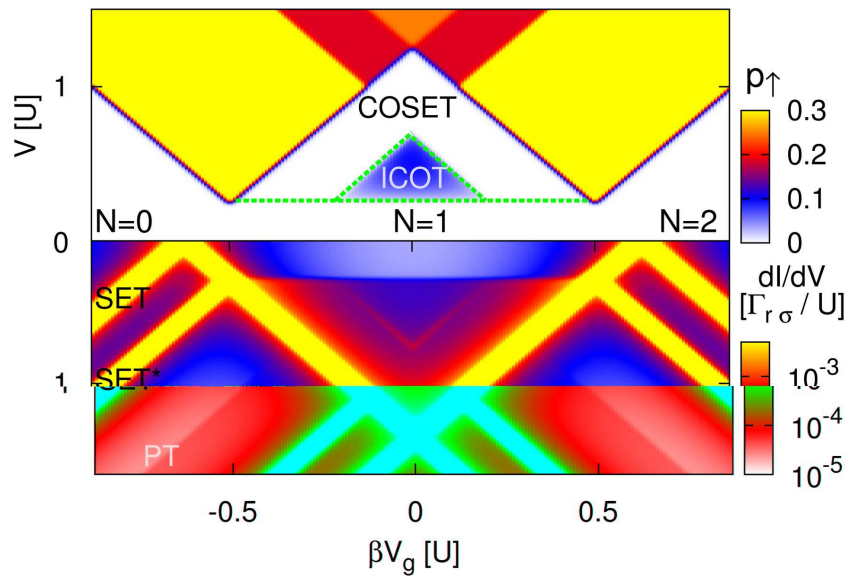}
  \end{center}
  \caption{
    (Color online)
    Occupation of excited spin-state, $p_\uparrow$ (upper panel) and differential
    conductance (lower panel) for the single-level Anderson model,
    plotted as a function of bias voltage $V$ and gate voltage $\beta V_\mathrm{g}$,
    where $\beta$ is the gate coupling factor.
    The spin degeneracy is lifted by an applied magnetic field:
    $\varepsilon_{\uparrow} - \varepsilon_{\downarrow} = 50T$
    where $T$ is the electron temperature.
    The dot is symmetrically coupled to the left and right electrodes:
    $\Gamma_{\mathrm{L}} = \Gamma_{\mathrm{R}} = 10^{-2}T = 5 \times 10^{-5} U$.
    The horizontal green line indicates the inelastic cotunneling threshold, which equals the Zeeman energy.
    The skewed green lines indicate where the sequential relaxation of the spin-excited state becomes 
    possible (COSET).
  }
  \label{fig:spectro}
\end{figure}
%%%%%%%%%%%%%%%%%%%%%%%%%%%%%%%%%%%%%%%%%%%%%%%%%%%

In figure~\ref{fig:spectro} we show exemplary results for the Anderson model of the calculations outlined above.
A complete overview of non-linear transport resonances is obtained  when plotting $dI/dV$
(quantifying changes in the current as new transport processes become energetically allowed)
as function of the static applied bias $V$ (which drives the current)
 and gate voltage $V_{\mathrm{g}}$ (which uniformly shifts all energy levels $\epsilon_\sigma$  without changing the quantum states)~\footnote{Here one assumes to a good first approximation that the bias and gate electric fields are spatially uniform. Corrections to this picture have been calculated in~\cite{Kaasbjerg08} for molecular junctions.
}.
In the lower half of the panel we plot  the differential conductance $dI/dV$,
whereas in the upper half of the panel the occupation $p_{\uparrow}$ of the spin-excited state of the quantum dot is shown.
This figure illustrates the insight provided by transport theory, linking the measured tunnel current to what goes on in the quantum dot.
To emphasize this correspondence of the two panels, the bias axes in the lower panel is mirrored with respect to
the upper.

To the far left and right of the plot the dot is in the $N=0$ and $N=2$ state respectively around zero bias.
Moving towards the center from there, $dI/dV$ shows a peak along the first encountered yellow skewed lines (marked ``SET'') where the condition
$\mu_\alpha = \varepsilon_\downarrow$ and $\mu_\alpha = \varepsilon_\uparrow+U$ are met, respectively.
Beyond this line a spin $\downarrow$ electron can now enter or leave the dot in a sequential or single-electron tunnel (SET) process which is of first order in $\Gamma$.
In this way the $N=1$ ground state $s=\downarrow$ can be reached starting from $N=0$ and $N=2$ respectively.
Once beyond the next skew yellow line (marked ``SET*''), defined by
$\mu_\alpha = \varepsilon_\uparrow$ and
$\mu_\alpha = \varepsilon_\downarrow+U$,
respectively,
an electron with spin $\uparrow$ can tunnel onto the dot.
This first order process makes the $N=1$ excited state $s=\uparrow$ accessible, as the upper panel clearly shows.

Moving further towards the center, there is a large triangular region of the $dI/dV$ map where only the $N=1$ ground state is accessible:
sequential tunneling processes out of this state are suppressed to first order since $U\gg T$, which is referred to as Coulomb blockade.
However, a second order process, referred to as inelastic cotunneling, allows the excited state to become occupied~\cite{DeFranceschi01,Lehmann05}
when the bias exceeds the excitation energy for a spin flip,
$\mu_{\mathrm{L}}-\mu_{\mathrm{R}} = V > \varepsilon_\uparrow-\varepsilon_\downarrow$.
This defines a gate-voltage independent line (horizontal),
 reflecting the fact that the charge state is left unaltered by the spin-fluctuation process which flips the spin.
Below this onset voltage only elastic cotunneling is possible, giving a smaller conductance and current.
In the small triangular region marked ``ICOT'' in the upper panel relaxation of the excited state is possible only by inelastic cotunneling.
Above this, in the region marked ``COSET'', the spin excited state, populated by slow inelastic excitation, can relax by  a much faster sequential relaxation process. It therefore is depleted again and the $dI/dV$ shows a small peak as signature of this cotunneling-assisted sequential tunneling (COSET)~\cite{Golovach04}.

The above processes are well-known and have been experimentally observed in various types of quantum dots, see, e.g., ~\cite{DeFranceschi01,Schleser05,Huettel09} and the reviews~\cite{Natelson06,vanderZant06,Biercuk08rev,Osorio08rev}.
However, only recently, it was shown that for this elementary model yet another resonance exists in second order in $\Gamma$~\cite{Leijnse09a}
by taking into account all many-body states and all contributions to the transport kernel $\Sigma$.
This resonance is related to \emph{electron pair tunneling}: it shows up along a skewed line in the $dI/dV$ map marked ``PT'',
 the change in the contrast indicating a step in the conductance as function of bias $V$.
The resonance condition for this is
$2\mu_\alpha = \sum_\sigma \varepsilon_\sigma+ U$,
indicating that the energy is paid (gained) by two electrons coming (going) from reservoir $\alpha$.
This electron pair can be added to (extracted from) the dot in one coherent process, thereby changing the charge from $N=0$ to $N=2$ (or vice versa).
The electro-chemical potential $\mu_\alpha = \frac{1}{2} \left(\sum_\sigma \varepsilon_\sigma+ U \right)$ at which this happens equals the \emph{average} to the above mentioned SET resonances, i.e., due to quantum charge fluctuations the Coulomb energy paid (gained) per electron is reduced by $\frac{1}{2}$.

Finally, we note that the resonance positions of the sequential tunneling peaks are slightly shifted with respect to the above mentioned values $\mu_\alpha = \epsilon_\downarrow,\epsilon_\uparrow+U$ due to the second order energy level renormalisation which is consistently accounted for.
Although this may seem unimportant here, in section~\ref{sec:external_fields} it will be shown that in time-dependent transport this level renormalisation effect may actually \emph{generate} and dominate the full current.

\subsection{Complex excitation spectra: vibrations and spin-orbit splitting}
\label{sec:complex}
%%%%%%%%%%%%%%%%%%%%%%%%%%%%%%%%%%%%%%%%%%%%%%%%%%%
\begin{figure}
  \begin{center}
    \includegraphics[width=1.0\linewidth]{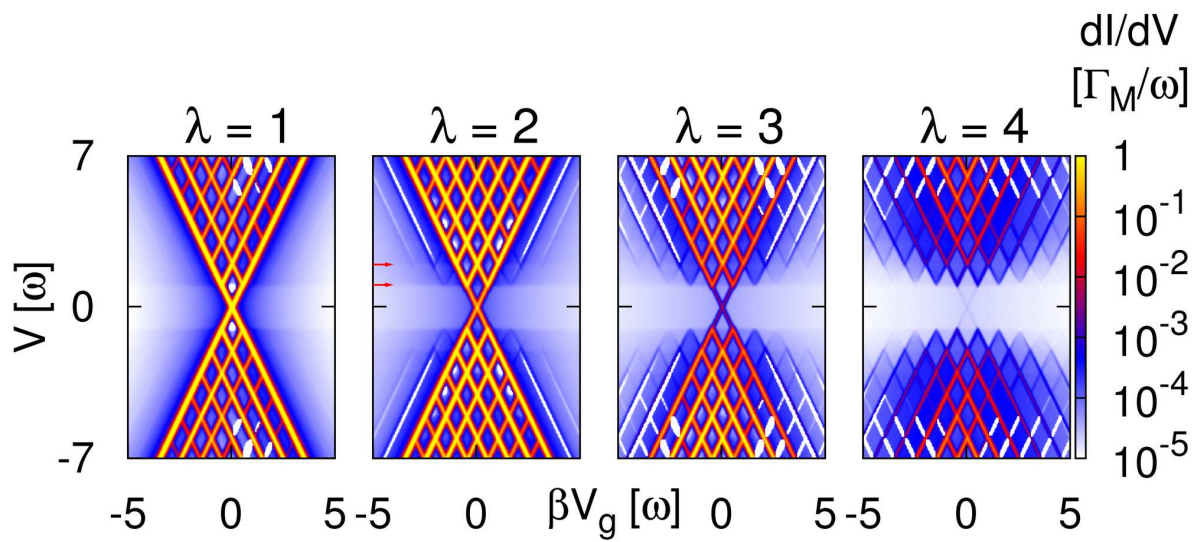}
  \end{center}
  \caption{
    (Color online)
    Differential conductance map vs. applied bias $V$ and gate voltage $V_g$ ($\beta=$ gate coupling factor)
    for a molecular level coupled to a vibrational mode with frequency $\omega$
    and strong local interaction $U \gg T,V,\lambda$~\cite{Leijnse08a}.
    With increasing electron vibration coupling $\lambda$, the inelastic cotunneling steps 
    (red arrows in the $\lambda=2$ figure indicate 1 and 2 phonon absorption)
    become relatively important with respect to the suppressed SET processes.
    The white lines / regions correspond to negative $dI/dV$, which cannot be plotted due to the 
    logarithmic color scale.
    Strikingly, the current thus displays a peak in the Coulomb blockade regime, which is very uncommon.
    In all plots we have chosen $\omega=40 T = 10^{4}\Gamma$ and conduction bandwidth $D=250\omega$.
    The conductance is scaled to $\Gamma_M$, the maximal sequential tunneling rate, i.e., $\Gamma$ times the 
    maximal Franck-Condon overlap factor (which is less than 1).
  }
  \label{fig:fc}
\end{figure}
%%%%%%%%%%%%%%%%%%%%%%%%%%%%%%%%%%%%%%%%%%%%%%%%%%%
Molecular quantum dots are clearly more complex than the above considered model, for instance by their ability to mechanically vibrate.
One important general aspect of this complexity is that the only conserved quantity in a tunnel process is 
the electron number~\footnote{Conserved here refers to the system of reservoirs and dot, including their interaction}.
In contrast, the quantum numbers of, e.g., various vibrational modes of molecular quantum dot may change when an electron tunnels onto it.
This implies that the transport theory has to account explicitly for the density matrix off-diagonal elements between different excited states with the same charge number $N$.
As is well known, this is important already in first order in $\Gamma$ when there are degeneracies of the energy levels $E_s$ in the same 
charge ($N$) state on the scale set by the tunneling~\cite{Braun04set,Wunsch05,Darau09}.
In the case where there are no such degeneracies only the probabilities matter to order $\Gamma$.
However, it was shown only recently, that even in this simplest case  this is no longer true in second (or higher) order in $\Gamma$ and \Eref{eq:stationary_master} can no longer be used.
Instead, one can derive from~\eref{eq_general_master} an effective stationary state master equation for the probabilities of the same form as~\eref{eq:stationary_master}, but with an effective kernel which contains important corrections from the off-diagonal elements~\cite{Leijnse08a,Koller10}.
An example where these corrections are found to be very important is the Franck-Condon effect.
This effect arises in its most elementary form in the Anderson-Holstein model, where the dot is described by
an Anderson level (discussed in section~\ref{sec:spectro}) plus a single vibrational mode with frequency $\omega$:
\begin{eqnarray}
  \label{eq:ham-fc}
  H_{\mathrm{D}} & = & \sum_\sigma \varepsilon_\sigma n_\sigma + U n_\uparrow n_\downarrow + \lambda \sqrt{2} Q \sum_\sigma n_\sigma + \frac{\omega}{2} (P^2 + Q^2)
  .
\end{eqnarray}
Here $Q=(b^\dagger+b)/\sqrt{2}$ is the vibrational coordinate normalized to the zero-point motion and 
$P=\rmi (b^\dagger-b)/\sqrt{2}$ the conjugate momentum.
Due to the linear coupling term $\propto \lambda$ the vibrational mode is displaced when charging the molecule.
For $\lambda > 1$ this causes a suppression of the Franck-Condon overlap of ground-state vibrational wave functions for two subsequent molecular charges $N$.
As a result the tunnel rates calculated to lowest order in $\Gamma$ are suppressed and the sequential tunneling current is blocked,
even for gate voltages close to the resonance~\cite{Boese01,McCarthy03,Flensberg03},
an effect also called ``Franck-Condon blockade''~\cite{Koch04b}.
This implies that second order processes become of crucial importance even at resonance~\cite{Koch06}.
The general approach developed in~\cite{Leijnse08a} consistently accounts for all second order effects,
i.e., not just cotunneling rates as in~\cite{Koch06}.
In figure~\ref{fig:fc} we show the differential conductance calculated to second order where the vibrational mode is shifted by several multiples of its zero-point amplitude.
Whereas for $\lambda=1$ the standard Coulomb blockade picture is still identifiable,
for $\lambda=4$ it is drastically altered:
inelastic cotunneling (threshold at $\mu_{\mathrm{L}}-\mu_{\mathrm{R}} = \omega, 2\omega, \ldots$) and COSET~\cite{Lueffe08} resonances proliferate in the transport spectrum.
Experimentally, vibration assisted transport in three terminal molecular devices has been reported~\cite{Park00,Pasupathy04,Osorio07a}.
Significant Franck-Condon blockade has been observed in suspended carbon nanotubes and compared with theory in~\cite{Leturcq09}.
Other theoretical studies have studied the effect on current noise~\cite{Koch04b,Koch04c}, the Kondo effect~\cite{Paaske05,Kikoin06}, the influence of mechanical dissipation (finite Q-factor)~\cite{Braig03a,Braig04b} or the lack thereof \cite{Koch05a}, vibrational anharmonicity~\cite{Wegewijs05,Koch05b} and interference effects~\cite{Wegewijs05,Donarini06}.

All above works rely on the Born-Oppenheimer separation of nuclear and electronic motion on the molecule.
Transport effects signaling the breakdown of this separation, going under the generic name of (pseudo-) Jahn-Teller dynamics, have also been calculated~\cite{Kaat05,Schultz07a,Reckermann08b}.
A very recent STM experiment~\cite{Repp10} has confirmed these transport signatures of the breakdown of the Born-Oppenheimer approximation.
In this work SET resonances were measured as function of a molecular parameter, as suggested in~\cite{Reckermann08b}, in this case by measuring many individual molecular wires of various lengths.

Another class of examples of quantum dots with complex excitations are magnetic molecules.
Here the spin-orbit interaction results in violation of  spin-selection rules for electron tunneling
and the same remarks as above apply.
In particular, single molecule magnets, reviewed in~\cite{Gatteschi06}, offer a basic example
and have been of interest in view of quantum computing~\cite{Leuenberger00,Lehmann07b}.
In such molecules the ground state has a large spin value $S>1/2$ due to strong exchange coupling of the spins of a few magnetic atoms.
This multiplet is however split by strong spin-orbit perturbations.
In many cases an accurate model for a given charge state $N$ is
\begin{eqnarray}
  \label{eq:spinham}
  H_{\mathrm{D}}^{(N)} & = & -D_N S_z^2 + E_N \left( S_x^2-S_y^2 \right)
  ,
\end{eqnarray}
where the uni-axial ($D_N$) and transverse ($E_N$) anisotropy parameters describe the splitting and mixing of spin states, respectively.
Recently, three-terminal measurements on magnetic molecules have been reported, including comparison with theory~\cite{Heersche06,Jo06,Grose08,Zyazin10}, see for a review~\cite{Bogani08}.
Calculations of the sequential~\cite{Romeike06b,Timm06,Elste06,Timm07,Gonzalez07,Lehmann07a}  
and  cotunneling~\cite{Elste07,Misiorny07a,Misiorny07b,Misiorny09,Misiorny10} 
regimes for  models including the anisotropy have identified distinctive transport signatures and possibly useful effects.
An important goal is to gain control over the spin of a single molecule by electric means~\cite{Osorio07b,Osorio10}.
Recently, the magnetic anisotropy splittings were measured in a three terminal junction, similar to two terminal STM measurements~\cite{Otte08,Otte09}.
Importantly, the magnetic field evolution of inelastic cotunneling~\cite{Zyazin10} was measured in different charge states.
It was found that the magnetic $D_N$ parameter depends strongly on the charge state $N$, which can be electrically controlled with the gate voltage.

Beyond the issue of control over single-molecule magnetism, is the understanding of coupled molecular magnets or magnetic centers.
Apart from spin-blockade effects, expected for large spin molecular ground states~\cite{Romeike07a},
the effect of spin-orbit interaction on the exchange coupling between two magnetic centers inside a magnetic \emph{double quantum dot} is of interest.
For instance, the spin-orbit induced Dzyaloshinskii-Moriya or antisymmetric exchange interaction was shown to give rise to a characteristic dependence of the transport on an externally applied magnetic field and the polarizations of magnetic electrodes~\cite{Herzog10}.

Higher order tunneling effects for magnetically anisotropic molecular quantum dots described by~(\ref{eq:spinham}) have attracted attention~\cite{Leuenberger06,Roosen08,Zitko08} after the prediction of a novel anisotropic pseudo spin Kondo effect~\cite{Romeike06a}.
The spin-orbit induced anisotropy in~\eref{eq:spinham} not only raises a barrier which  opposes spin-fluctuations (due to $D_N$) (section~\ref{sec:fluct_linear}),
but it also generates intrinsic spin-tunneling (due to $E_N$).
This interplay leads to interesting anisotropic effective exchange interaction with electrons transported through a molecular magnet, as will be discussed in section~\ref{sec:fluct_nonlinear}.

Besides their own interesting transport signatures, spin and vibrations can also interact or influence each other in transport.
For instance, a mechanism of \emph{vibration-induced} spin-blockade of transport was predicted for a mixed-valence dimer molecular transistor (``double dot'')~\cite{Reckermann09a,Wegewijs09a}.

Finally, we mention that level spectrum complexity may also deceive one when analyzing experimental data.
A particularly clear example is offered by recent transport data on carbon-nanotube ``peapods'', i.e. fullerene molecules in a host nanotube~\cite{Eliasen10}.
Here weakly gate-dependent $dI/dV$ peaks were observed which are strongly reminiscent of inelastic cotunneling (second order in $\Gamma$, see section~\ref{sec:spectro}).
However, detailed transport calculations have  conclusively shown that the complex measured spectrum can be assigned to sequential tunneling processes only, i.e., of \emph{first order} in $\Gamma$.
This deceptive case occurs because the two coupled quantum dots, the host nanotube and the fullerene ``impurities'' have a very different gate voltage dependence.
Fortunately, the significant hybridization of the two dots gives a characteristic interference effect in the first order current, distinguishing it clearly form second order effects.
In view of applications it is interesting that the analysis  in~\cite{Eliasen10} indicates that the charge state of the fullerene impurities is electrically tunable, independently of that of the host carbon nanotube.

\subsection{Outlook}
In summary, we have illustrated that general methods for spectroscopy calculations for molecular quantum dot models are essential for the experimental characterization and ultimately the design of nanoscale quantum dot devices.
The complex spectra due to various strong many-body interactions play a crucial but complicating role,
at the same time however, enabling interesting new possibilities.
An exciting future direction important for further progress is to efficiently account for renormalisation effects in complex models,
 in particular as outlined in section~\ref{sec:fluct_nonlinear}.
Also, the approach outlined here can be extended to other transport quantities, such as spin and heat~\cite{Segal06}.
The molecule and junction models used here can be parametrized based on atomistic electronic structure calculations, as for instance, in~\cite{Hettler03,Seldenthuis08,Darau09}.
Pursuing this further is another important step in understanding molecular quantum dots.

%%%%%%%%%%%%%%%%%%%%%%%%%%%%%%%%%%%%%%%%%%%%%%%%%%
\section{Adiabatic transport  through a quantum dot due to time-dependent fields}
\label{sec:external_fields}
%%%%%%%%%%%%%%%%%%%%%%%%%%%%%%%%%%%%%%%%%%%%%%%%%%%

Charge transport through quantum dots is not necessarily due to an applied static bias voltage: in the absence of a bias, a current can be obtained by the time-dependent modulation of fields, applied externally to a mesoscopic device. This is of interest for possible future applications in nanoelectronics, where the repeated and fast operation of a device is of interest.  Indeed the mesoscopic analogue of a capacitor has been realized~\cite{Gabelli06,Feve07} serving as a fast coherent electron source and charge pumping~\cite{Blumenthal07,Fuhrer07,Buitelaar08,Ebbecke08} is a candidate for the realization of a quantum current standard, to name some examples. From a fundamental point of view the nonequilibrium created by time-dependent electric fields provides enhanced insight into the quantum properties of a nanoscale device, which are not accessible with equilibrium methods. If the variation of the fields is slow with respect to the internal time-scales of the system itself, the modulation is called \textit{adiabatic}. The system then remains in a quasi-equilibrium state, meaning that  the system is not brought to an excited state by the time-dependent modulation~\footnote{The slow variation of the externally applied fields becomes experimentally particularly important for delicate systems as molecules, see section \ref{sec:molecules}, where problems related to heating, surface excitations (plasmons) and laser-induced chemical reactions (photo-bleaching) should
be avoided.}.

\begin{figure}
\begin{center}
\includegraphics[width=6cm]{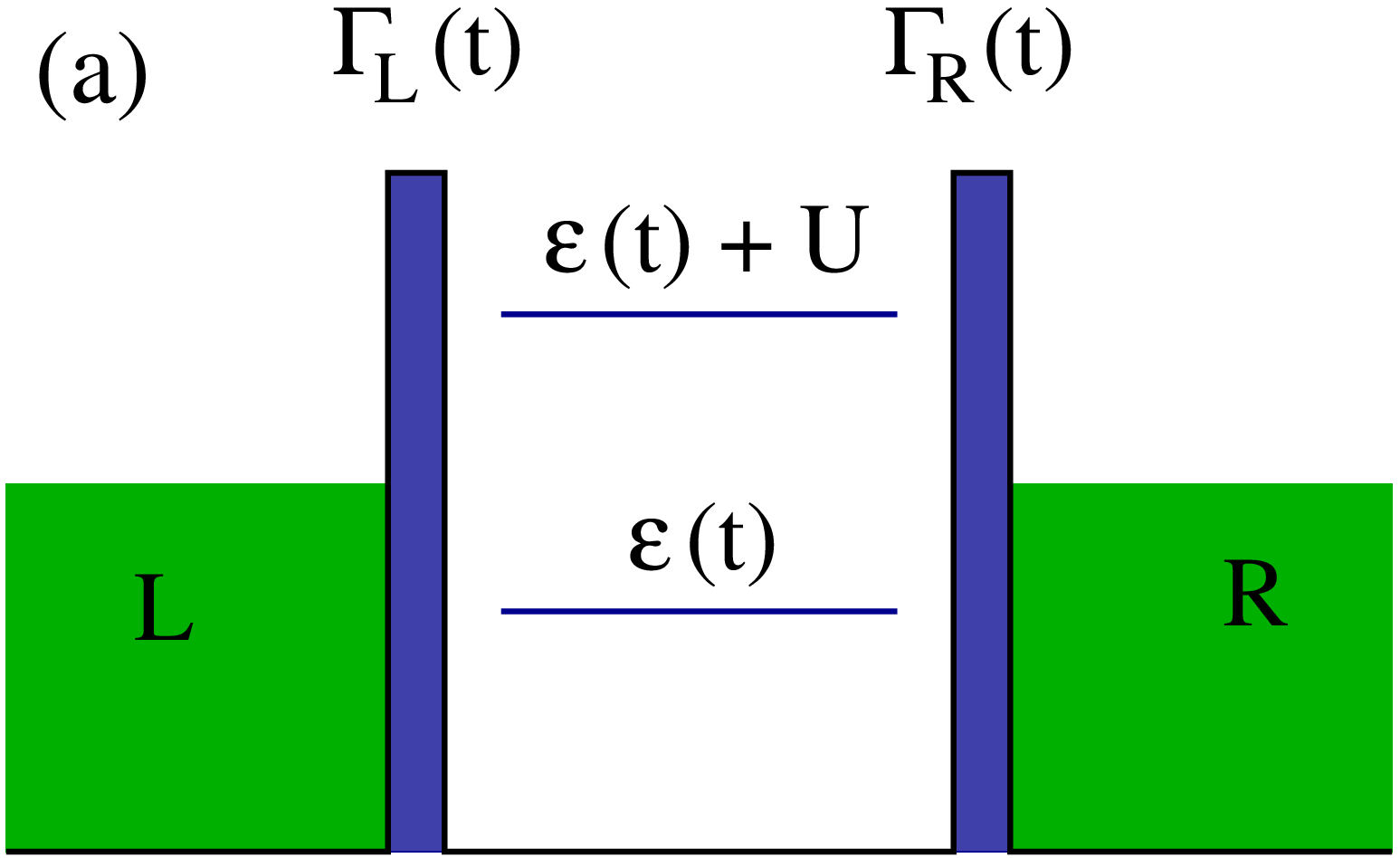}
\hspace{1cm}
\includegraphics[width=4cm]{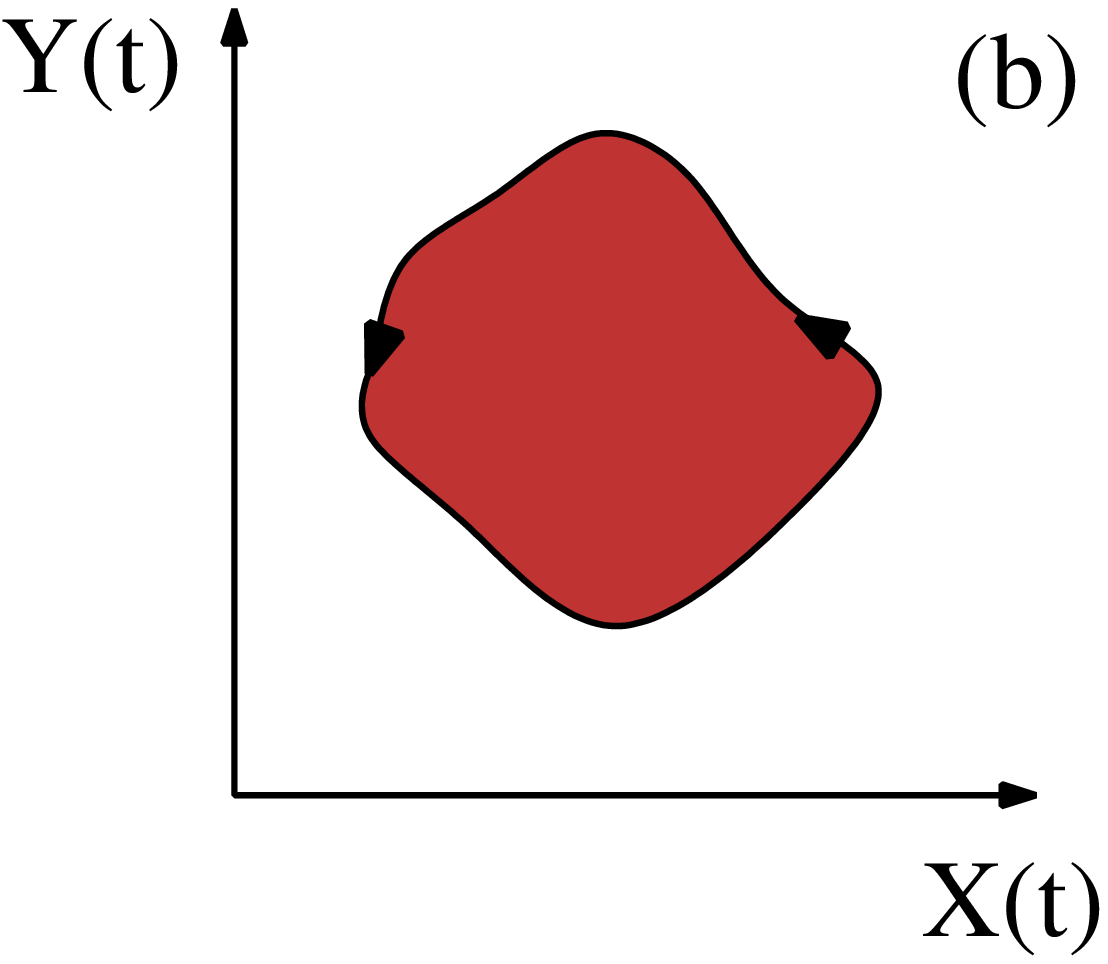}
\end{center}
\caption{(Color online) (a) Sketch of a single-level quantum dot with time-dependent parameters attached to two reservoirs. (b) Two of the system parameters are time dependent as indicated in (a) and perform a closed cycle in parameter space.}
\label{fig_model}
\end{figure}

A prominent example for transport in the absence of a bias voltage is \textit{adiabatic pumping}: the cyclic variation of at least two of the system parameters, see figure~\ref{fig_model}, leads to a dc charge transfer through the quantum device.  In this case the pumped charge 
does surprisingly not depend on the detailed time evolution of the pumping cycle; this can be shown by formulating it in terms of geometric phases~\cite{Avron00,Makhlin01,Zhou03}.
Theoretically, adiabatic pumping has been extensively studied in the non-interacting regime~\cite{Brouwer98,Zhou99,Levinson00,Aleiner00,Levitov01,Moskalets01,Moskalets02}. In this case a scattering matrix approach is convenient~\cite{Buettiker94,Brouwer98} and has been broadly used.
While indeed Coulomb interactions are of minor importance in some setups~\cite{Gabelli06,Feve07,Switkes99}, they play an important role in many quantum-dot systems and lead to a variety of interesting results.
Several approaches have been used to address the problem of adiabatic pumping through interacting
quantum dots in different regimes, including weak Coulomb interaction~\cite{Aleiner98,Brouwer05}, pumping in Luttinger liquids~\cite{Citro03}, and a quite general approach using nonequilibrium Green's functions~\cite{Splettstoesser05,Sela06,Fioretto08}. The latter works are particularly appropriate for the study of weak interaction or for the calculation of spin and charge pumping in the Kondo regime~\cite{Sela06}  and the role of elastic and inelastic scattering processes has been studied in detail~\cite{Fioretto08}. The treatment of the tunnel coupling is non-perturbative in these works and they are in this sense complementary to the results presented in the following. 

It turns out that the interplay of Coulomb interaction and coherent tunneling between a quantum dot and leads, leading to an effective renormalisation of the energy level, can be at the origin of the pumping mechanism. The result is that this energy renormalisation, which is a pure Coulomb interaction effect vanishing when Coulomb interactions are negligible, is accessible via the pumped charge.

Here we present results in the regime in which the coupling between the dot and the leads is 
moderate; this means that the energy scale $\Gamma$ approaches temperature $T$ but is still
smaller than $T$ so that charge fluctuations start to become important. In contrast, the energy scale 
$T_\mathrm{K}$ of the Kondo temperature setting the
scale for the onset of spin fluctuations, is taken to be much smaller than $\Gamma$ and $T$.
To approach the effect described above we use a perturbative expansion in the tunnel coupling 
between the dot and the leads up to second order in $\Gamma$, which is appropriate in this regime.
The aim is the understanding of the influence of finite and arbitrary Coulomb interaction on the pumping characteristics and the identification of the physical origin of the various contributions to the pumped charge. 
In order to achieve this, a diagrammatic real-time technique~\cite{Koenig96a,Koenig96b,Schoeller97hab,Koenig99}, which  has been developed to describe non-equilibrium DC transport through an interacting quantum dot, 
see section \eref{sec:molecules}, was extended for the treatment of adiabatically time-dependent fields~\cite{Splettstoesser06}. 
%%%%%%%%%%%%%%%%%%%%%%%%%%%%%%%%%%%%%%%%%%%%%%%%%%%
%%%%%%%%%%%%%%%%%%%%%%%%%%%%%%%%%%%%%%%%%%%%%%%%%%%
\subsection{Adiabatic expansion}

Even though the approach is not restricted to specific systems, here the focus is put on a single-level quantum dot as shown in figure \ref{fig_model}. The quantum dot has a time-dependent level $\varepsilon(t)=\bar{\varepsilon}+\delta\varepsilon(t)$ and is coupled to leads $\alpha=\mathrm{L,R}$ with time-dependent tunnel amplitudes $T_\alpha(t)$. The strength of the tunnel coupling is characterized by the time-dependent quantity, $\Gamma_\alpha(t)=2\pi\rho_\alpha T_\alpha^*(t)T_\alpha(t)=\bar{\Gamma}_\alpha+\delta\Gamma_\alpha(t)$, with the lead's constant density of states, $\rho_\alpha$. As described in section~\ref{sec:molecules}, the eigenstates of the decoupled system are given by $s=\left\{0,\uparrow,\downarrow,\mathrm{d}\right\}$. Tracing out the lead degrees of freedom the system is described by its reduced density matrix, where here only the occupation probabilities $\bi{p}=  (p_0,p_\uparrow,p_\downarrow,p_{\mathrm{d}})^{\mathrm{T}}$ are of interest. 
Their time-evolution in the long-time limit, $t_0\rightarrow-\infty$, is fully described by the generalized Master equation,~\Eref{eq_general_master}.
%\begin{equation}
%\label{eq_general_master}
%	\frac{d}{dt}\bi{p}\left(t\right)=-i\int_{-\infty}^{t}dt'\boldsymbol{\Sigma}
%	\left(t,t'\right)\bi{p}\left(t'\right)\, .
%\end{equation}
The goal being the description of slowly varying system parameters, it is useful to perform an adiabatic expansion~\cite{Splettstoesser06}: the frequency of the variation therefore has to be much smaller than the dwell time of electrons in the system, $\Omega\ll\Gamma$. 
The effect of the system parameters varying slowly in time is that the density matrix of the system at time $t$ is not only described by the instantaneous values of the parameters, but it lags a bit behind the parameter change. Therefore an expansion around this time $t$ suggests itself. To obtain such an expansion as a first step a Taylor expansion of $\bi{p}(t')$ is performed around $t$ up to
linear order in the integrand on the right hand side of ~\Eref{eq_general_master}. This is related to the fact that memory effects of the kernel have to be taken into account. Furthermore, an adiabatic expansion of the kernel  $\boldsymbol{\Sigma}\left(t,t'\right)$ itself is performed.
The zeroth-order term, $\boldsymbol{\Sigma}^{(i)}_{t}(t-t')$, is indicated with the 
superscript $(i)$ for {\em instantaneous}. The subscript $t$ to emphasize 
that the system parameters $X(\tau) \rightarrow X(t)$ are frozen at time $t$, i.e., the functional dependence on $X(\tau)$ is replaced by $X(\tau) \rightarrow  X(t)$.
The first-order term is obtained by linearizing the time dependence of all 
parameters $X(\tau)$ with respect to the final time $t$, i.e., 
$X(\tau) \rightarrow X(t)+ (\tau-t) \frac{d}{d\tau} X(\tau)|_{\tau=t}$, and 
retaining only linear terms in time derivatives.
This linear correction to the kernel is indicated by 
the superscript $(a)$ for {\em adiabatic}. Finally, it is necessary to perform an adiabatic expansion for the occupation probabilities in the dot,
\begin{eqnarray} 
\label{eq_adexp}
	\bi{p}(t) & \rightarrow & \bi{p}^{(i)}_{t}+\bi{p}^{(a)}_{t}
\\
  	\boldsymbol{\Sigma}(t,t')  & \rightarrow & \boldsymbol{\Sigma}^{(i)}_{t}(t-t') + 
  	\boldsymbol{\Sigma}^{(a)}_{t}(t-t') 
	\end{eqnarray}
The instantaneous probabilities $\bi{p}^{(i)}_{t}$ are the solution 
of the time-independent problem with all parameter values fixed at time $t$.
They are obtained by solving the stationary Master equation, 
 \Eref{eq:stationary_master}, with all parameters replaced by 
 time-dependent parameters evaluated at time $t$. Once the instantaneous
probabilities $\bi{p}^{(i)}_t$ are known, the adiabatic corrections 
$\bi{p}^{(a)}_t$ are
found from the Master equation in first order in $\Omega$ obtained from the expansion described above.
Similarly an equation for the current 
%\begin{eqnarray}
%\label{eq_current}
%	I_{\mathrm{L}}\left(t\right)=-ie\int_{-\infty}^{t}dt'\bi{e}^{\mathrm{T}}
%	\boldsymbol{\Sigma}^{\mathrm{L}}\left(t,t'\right)\bi{p}\left(t'\right),
%\end{eqnarray}
can be developed, see equation~\eref{eq_current}, and adiabatically expanded, 
where a current kernel $\boldsymbol{\Sigma}^{\mathrm{L}}$ takes account for the amount and the direction of electron transfer. 

On top of the adiabatic expansion a perturbative expansion in the tunnel coupling strength $\Gamma$ is performed
up to second order, allowing the actual evaluation of the kernel using diagrammatic rules as elaborated in Refs.~\cite{Koenig96a,Koenig96b,Schoeller97hab,Koenig99,Splettstoesser06}, see section \ref{sec:molecules}. 
This perturbative expansion is valid in the regime of moderate couplings with respect to temperature, 
i.e. $\Gamma\lesssim T$.

%%%%%%%%%%%%%%%%%%%%%%%%%%%%%%%%%%%%%%%%%%%%%%%%%%%
\begin{figure}
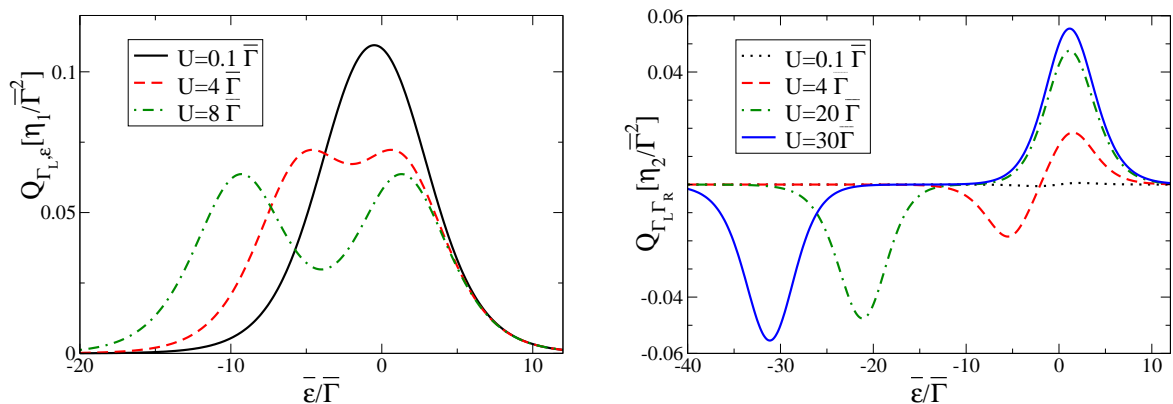

\begin{center}
\includegraphics[width=7.4cm]{pumped_lowest.eps}
\hspace{0.5cm}
\includegraphics[width=7.3cm]{pumped_first.eps}
\end{center}
\caption{(Color online) (a) Pumped charge due to a parameter variation of the dot level and of one of the barriers as a function of the average dot level. The dominant contribution to the  pumped charge is due to first order tunneling processes.  (b) Pumped charge due to the pure variation of the barrier strengths as a function of the average dot level. The dominant contribution to the  pumped charge is due to \textit{second} order tunneling processes only. In both plots the temperature equals $T=2\Gamma$.
\label{fig_pumped_charge}}
\end{figure}
%%%%%%%%%%%%%%%%%%%%%%%%%%%%%%%%%%%%%%%%%%%%%%%%%%%
\subsection{Pumping current}

We are interested in the time-resolved pumping current as well as the  charge pumped through the dot per pumping cycle. The pumped charge, $Q=\int_0^{\mathcal{T}} I_{\mathrm{L}}(t) dt$, in the bilinear response regime, i.e. for modulations with small amplitudes, is proportional to the area spanned in parameter space, see~\Fig{fig_model}. As a first step the time-resolved  pumping current is calculated in lowest order in the tunnel coupling. It is straightforward to relate the pumped current to the  dynamics of the average instantaneous charge of the dot 
$\langle n\rangle^{(i)}_t$ and one finds
\begin{eqnarray}
  	\label{eq_first_current}
  	I_{\mathrm{L}}^{(0)}\left(t\right) & = & 
              -\frac{\Gamma_{\mathrm{L}}(t)}{\Gamma(t)}
  	\frac{d}{dt}\langle n\rangle^{(i,0)}_t \, , 
\end{eqnarray}
where $\langle n\rangle^{(i,0)}_t$ denotes the instantaneous occupation in lowest order in $\Gamma$.
This suggests the following interpretation. As the dot occupation in lowest order, $\langle n\rangle^{(i,0)}_t$, is changed in time by varying the pumping parameters  the charge moves in and out of  
the quantum dot generating a current from/into the leads. Note that one of the pumping parameters must be the level position since $\langle n\rangle^{(i,0)}_t$ is given by Boltzmann distributions and is therefore 
independent of the tunnel-coupling strengths. This means that $\frac{d}{dt}\langle n\rangle^{(i,0)}_t=0$ if the level position is constant. The charge moving in and out of the quantum dot   is divided to the left and to the right, depending on the time-dependent relative tunnel couplings $\Gamma_{\alpha}(t)/\Gamma(t)$. 

The next step is to calculate the next-to-leading order correction to the lowest order result shown in ~\Eref{eq_first_current}. Also the next-to-leading order can be written in a compact way
\begin{equation}
	I_{\mathrm{L}}^{(1)}\left(t\right) =  -\left\{\frac{d}{dt}
	\left(\langle n\rangle^{(i,\mathrm{broad,L})}_t
	\right)+\frac{\Gamma_{\mathrm{L}}(t)}{\Gamma(t)}\frac{d}{dt}\langle 
	n\rangle^{(i,\mathrm{ren})}_t
	\right\}. 
\label{eq_second_current}
\end{equation}  
The first term of ~\Eref{eq_second_current} contains the contribution
due to the correction of the average dot occupation induced by 
lifetime-broadening. It contains a total time derivative, 
and as parameters are periodically changing in time, it will not lead to a
net pumped charge after the full pumping cycle. 
The second term has the same structure as the zeroth-order contribution,
~\Eref{eq_first_current}.
It can be understood as the correction term 
induced by the combined influence of charge fluctuations and 
correlation effects giving
rise to a renormalisation of the level position, 
$\varepsilon(t)  \rightarrow  \varepsilon(t)+ \sigma\left(\varepsilon(t),\Gamma(t),U\right)$.
The level renormalisation, $\sigma\left(\varepsilon,\Gamma,U\right)$, is positive when the level is in the vicinity of the Fermi energy of the leads and negative for $\varepsilon+U$ being close 
to the Fermi energy. The sign of the renormalisation thus indicates if transitions take place between empty and singly occupied dot or between single and doubly occupied dot. The energy level renormalisation may be time dependent via time-dependent 
tunnel couplings or a time-dependent level. The result is that a finite
charge can be pumped by means of level renormalisation. This is a pure Coulomb interaction effect, it 
vanishes for $U\rightarrow0$. 
We note that correction terms from the renormalisation of the tunneling couplings
do not contribute in first order in $\Gamma$.
For the DC current, different contributions from higher order processes are present at the same
time, which makes it challenging to identify them separately in an experiment.
For the adiabatically pumped charge, where correction terms associated with a 
renormalisation of the tunnel couplings and level-broadening effects vanish, 
the situation is distinctively different.
Studying adiabatic pumping is, therefore, a convenient tool to access the 
energy-level renormalisation.
This is most dramatic in the case when the zeroth-order pumped current is zero,
i.e., when pumping is done by changing both tunnel couplings.
In this case, the {\em dominant} contribution to the pumped charge is due to
time-dependent level renormalisation. These results suggest that adiabatic pumping can be used to \textit{directly} access the level renormalisation in quantum dots.

The results for the pumped charge per pumping cycle are shown in~\Fig{fig_pumped_charge}, where $\eta$ is the enclosed surface in parameter space. Modulating the level position and one of the tunnel couplings the pumped charge has a maximum contribution at the resonances, see~\Fig{fig_pumped_charge}(a). Importantly the contribution from sequential tunneling is dominant here. Figure~\ref{fig_pumped_charge}(b) shows the pumped charge obtained by modulating the two tunnel couplings. As described above, the pumped charge is due to level renormalisation and therefore vanishes for vanishing Coulomb interaction. The sign change between the two contributions at the resonances reflects the opposite sign of the level renormalisation for the 
two resonances.

The method described here is generally formulated and it is therefore extendable to a variety of systems in which Coulomb interaction is important and  to which time-dependent fields are applied. Charge and spin pumping through an interacting quantum dot has been studied in the presence of ferromagnetic leads~\cite{Splettstoesser08}. Pumping through one or two metallic interacting islands with a continuous density of states has been examined in~\cite{Winkler09}. In~\cite{Cavaliere09} the validity of the frequency regime has been extended to faster modulations and in ~\cite{Hiltscher10} the influence of quantum interference was studied on the pumped charge through one or two quantum dots embedded in an Aharonov-Bohm geometry. Finally this approach has been used to study the capacitive and the relaxation properties of a driven quantum dot figuring as a mesoscopic capacitor in ~\cite{Splettstoesser10}.

\subsection{Outlook}

In this section we have presented results on adiabatic pumping, where interesting effects due to quantum charge fluctuations and finite Coulomb interaction are revealed. The underlying method shown here uses an adiabatic expansion of the Master equation based on a real-time diagrammatic technique; this method is applicable whenever a mesoscopic system is exposed to a certain number of adiabatically  time-dependent fields. 
The intriguing results found for systems with a relatively simple spectrum together with the \textit{general} formulation of the adiabatic expansion of the Master equation, motivate further investigations. A challenging question is, e.g., to study  the impact of time-dependent fields on molecular devices with a complex energy spectrum, as introduced in section~\ref{sec:molecules}. Another interesting development concerns the combination of the formalism of adiabatic quantum pumping with renormalisation group methods, as described in sections \ref{sec:fluct_linear} and \ref{sec:fluct_nonlinear}, to describe the influence of time-dependent external fields in the regime of strong correlations and strong quantum fluctuations.

\section{Quantum fluctuations in linear response} 
\label{sec:fluct_linear}

In this section we will discuss the physics of strong quantum fluctuations
in combination with correlation effects in quantum dots. We will concentrate on
the linear response and static regime, the dependence on finite bias voltage
together with the time evolution will be discussed in section \ref{sec:fluct_nonlinear}.
One of the most prominent examples of the drastic effects of spin fluctuations
in quantum dots is the experimental observation of the Kondo effect \cite{Goldhaber98}. 
In bulk solids a small amount of magnetic impurities leads to an increased 
magnetic scattering 
of the electrons at low temperature, which results in an increased resistance.
In quantum dots, the mesoscopic realization of a single spin coupled to two leads
displays instead a zero-bias peak in the conductance \cite{Glazman88}. 
The particular advantage of quantum dots is that they allow for a very flexible tuning 
of the parameters and can easily be extended
to study the effects of more complex impurities, where  orbital Kondo, quantum critical and 
interference effects may arise. Moreover, additional environmental degrees of freedom
in presence of ferromagnetic or superconducting reservoirs 
coupled to the quantum dot as well as finite-size effects affect the 
transport properties. For spin fluctuations, the characteristic energy scale $T_{\rm c}$ is
given by the Kondo temperature $T_{\rm K}$, which, for the special case of a single-level dot
with Coulomb energy $U$ and tunneling coupling $\Gamma$, is given by 
$T_{\rm K}\sim\sqrt{U\Gamma}e^{-{\pi U\over 4\Gamma}}$. This scale is exponentially small
for large Coulomb energy and, therefore, the Kondo effect in quantum dots is only
visible for strongly coupled leads. In contrast, the importance of charge fluctuations 
is controlled by $\Gamma$ and signatures are already significant for $\Gamma\sim T$.
Already in the last section, we showed how renormalised level positions
can be identified in the adiabatically pumped charge. Experimentally, effects from 
charge fluctuations have been detected in transmission phase lapses through multi-level
quantum dots in Aharonov-Bohm geometries \cite{Yacoby95,Schuster97}. New light was shed on
this long-outstanding puzzle by the insight that the combined influence of broadening
and renormalisation effects induced by charge fluctuations is responsible for this effect
\cite{Karrasch07a,Karrasch07b}.

The analysis of the signatures of correlation and quantum fluctuation effects requires
appropriate methods for their theoretical description.
The importance of the development of analytical as well as numerical techniques for their
treatment constitutes therefore an important issue.
In recent years functional renormalisation-group (fRG) methods \cite{Salmhofer98} have been 
established as a new computational tool in the 
theory of interacting Fermi systems. These methods are particularly powerful in low dimensions.
The low-energy behavior usually described by an effective field theory can be computed 
for a concrete microscopic model by solving a differential flow with the energy scale
as the flow parameter. Thereby also the nonuniversal behavior at intermediate scales is obtained.
For applications to quantum dots and wires the fRG scheme turned out to be an
efficient approach for the description of the single-particle spectral properties and the linear transport 
through generic models in presence of moderate correlations. 
Arbitrary tunneling couplings can be considered, allowing to access the regime of strong coupling between the reservoirs and the quantum system.
Due to the flexibility of the microscopic modeling the whole parameter range can be easily 
explored, and more complex geometries of multi-level quantum dots, as realized in 
recent experiments, can be treated. 
In the following sections we will discuss applications of the fRG to the description of
strong spin and charge fluctuations in quantum dots, exemplified by the discussion of the
Kondo effect and transmission phase lapses in multi-level quantum dots.

\subsection{The Kondo effect in quantum dots}

The appearance of an upturn followed by a low-temperature saturation of the resistivity 
in metals containing diluted magnetic impurities was explained by Kondo in 1960 as an enhanced 
scattering due to the screening of the local impurity spins by the conduction electrons 
\cite{Kondo64}. 
The Fermi liquid nature of the ground state \cite{Nozieres74} and the various theoretical treatments, 
including Wilson's numerical renormalisation group (NRG) \cite{Wilson75}, make the Kondo problem one of the 
best understood many-body phenomenon in condensed matter physics, as well as a paradigm for correlated electron 
physics \cite{Hewson93}. 
In the late 90's, the prediction by Ng and Lee \cite{Ng88}, as well as 
Glazman and Raikh \cite{Glazman88} of the Kondo effect in quantum dots led to a revival of the subject, 
and has been observed in 2D GaAs/AlGaAs  heterostructure quantum dots 
\cite{Goldhaber98,Cronenwett98,Schmid98}, silicon MOSFET's, 
as well as in carbon nanotubes \cite{Nygard00} and contacted single molecules \cite{Park02,Liang02}.
Confinement and electrostatic gating enabled to investigate the Kondo effect in
an artificial Anderson impurity through electrical 
transport measurements, characterized by a unitary conductance $2\,\frac{e^2}{h}$ at zero temperature 
\cite{Pustilnik04rev}. 
Besides providing a paradigm for a variety of physical effects involving strong electronic 
correlations,
the Kondo effect in nanostructures allows for the realization of spintronic device constituents as 
spin filters or quantum gates.
The understanding and control of the transport of spin-polarized currents
are of fundamental importance for the advancement of semiconductor spintronic 
device technology \cite{Awschalom02}.

The coupling of a quantum dot with spin-degenerate levels and local
Coulomb interaction to metallic leads gives rise to Kondo
physics \cite{Hewson93}. A small quantum dot with large level spacing is described 
by the single-impurity Anderson model (SIAM) in the regime in which only
a single spin-degenerate level is relevant.
At low temperatures and for sufficiently high
barriers the local Coulomb repulsion $U$ leads to a broad resonance
plateau in the linear conductance $G$ of such a setup as a function of
a gate voltage $V_{\rm g}$ which linearly shifts the level
positions \cite{Tsvelick83,Glazman88,Ng88,Costi94,Gerland00}. It
replaces the Lorentzian of width $\Gamma$ found for noninteracting electrons.  On
resonance the dot is half-filled implying a local spin-$\frac{1}{2}$
degree of freedom responsible for the Kondo effect \cite{Hewson93}. 
In the limit of large $U\gg\Gamma$ the charge degrees of freedom can be integrated out and
the spin physics is described by the Kondo model, which will be discussed in 
section \ref{sec:fluct_nonlinear}.
For the SIAM the zero temperature conductance is proportional to the
one-particle spectral weight of the dot at the chemical
potential \cite{Meir92}. The appearance of the plateau of width $U$ in the
conductance is due to the pinning of the Kondo resonance in the
spectral function at the chemical potential for $-\frac{U}{2}\lesssim
V_{\rm g}\lesssim \frac{U}{2}$ (here $V_{\rm g}\equiv \epsilon+{U\over 2}=0$ 
corresponds to the half-filled dot case) \cite{Hewson93,Gerland00}. Kondo physics in transport through
quantum dots was confirmed experimentally \cite{Goldhaber98,vanderWiel03rev},
and theoretically using the Bethe ansatz \cite{Tsvelick83,Gerland00} and
the NRG
technique \cite{Wilson75,Krishnamurthy80}.  However, both methods can
hardly be used to study more complex setups.  In particular, the
extension of the NRG to more complex geometries beyond single-level
quantum-dot systems \cite{Costi94,Gerland00,kondo_theo5} is restricted
by the increasing computational effort with the
number of interacting degrees of freedom.  Alternative methods which
allow for a systematic investigation are therefore required.  Here the
fRG approach is proposed to study
low-temperature transport properties through mesoscopic systems with
local Coulomb correlations.

A particular challenge in the description of quantum dots is their
distinct behavior on different energy scales, and the appearance of
collective phenomena at new energy scales not manifest in the
underlying microscopic model.  An example of this is the Kondo effect
where the interplay of the localized electron spin on the dot and the
spins of the lead electrons leads to an exponentially small (in $U/\Gamma$)
scale $T_{\rm K}$. This diversity of scales cannot be captured by
straightforward perturbation theory.  One tool to cope with such
systems is the renormalisation group: by treating different energy
scales successively, one can often find an efficient description for
each one.  The fermionic fRG \cite{Salmhofer01} is formulated in terms of
an exact hierarchy of coupled flow equations for the vertex functions
(effective multi-particle interactions) as the energy scale is lowered.
The flow starts directly from the
microscopic model, thus including nonuniversal effects on higher
energy scales from the outset, in contrast to effective field theories
capturing only the asymptotic behavior.  As the cutoff scale is
lowered, fluctuations at lower energy scales are successively included
in the determination of the effective correlation functions.  This allows to
control infrared singularities and competing instabilities in an
unbiased way.  Truncations of the flow equation hierarchy and suitable
parametrizations of the frequency and momentum dependence of the
vertex functions lead to new approximation schemes, which are
devised for moderate renormalised interactions. A comparison with exact
results shows that the fRG is remarkably accurate even for sizeable
interactions.

\begin{figure}[]
\vspace{.5cm}
  \centering
  \center{\includegraphics[width=.6\textwidth]{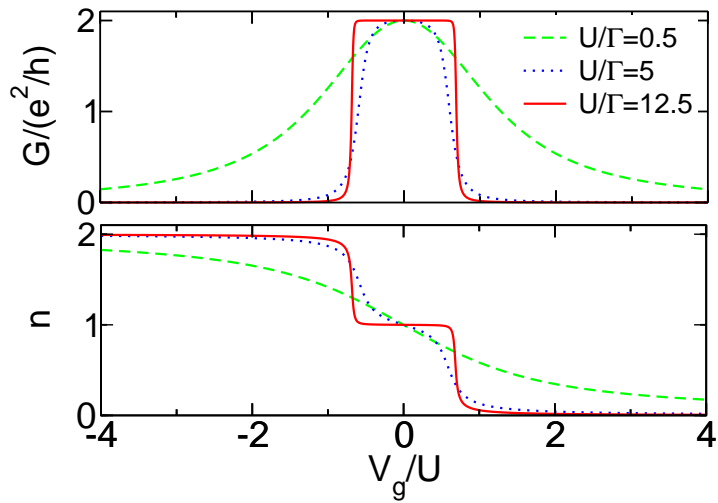}}
  \caption{(Color online) \emph{Upper panel:} conductance as 
    a function of gate voltage for different values of $U/\Gamma$.  \emph{Lower panel:} average number of electrons
    on the dot.}
  \label{fig:fig3}
\end{figure}

To exemplify the effect of correlations, we first focus on
a quantum dot with spin-degenerate levels. 
For simplicity only a single level with a local Coulomb repulsion $U$ described by the SIAM \cite{Hewson93} is considered, see figure~\ref{fig:skizze}.
The hybridization of the dot with the leads broadens
the levels on the dot by $\Gamma_{\alpha}= 2 \pi T_{\alpha}^2
\rho_{\alpha}$, where $\rho_{\alpha}$ is the local density of
states at the end of lead $\alpha=L,R$ assumed to be independent of frequency here.  
The energy level is determined by the gate voltage $V_{\rm g}$.
Integrating out the lead degrees of freedom,
the bare Green function of the dot  \cite{Hewson93} is
\begin{equation}
  {\mathcal{G}}_{0}(i\omega) = \frac{1}{i\omega-V_{\rm g}+i\frac{\Gamma}{2} \mathrm{sgn} (\omega)} \,,
\end{equation}
where $\Gamma = \Gamma_{\rm L} + \Gamma_{\rm R}$.
By solving the interacting many-body problem  a
 self-energy contribution $\Sigma(i\omega)$ is obtained dressing the bare propagator
through the Dyson equation.
The fRG is used to provide a self-energy describing the
physical properties of the $T=0$ linear conductance obtained 
by  $G(V_{\rm g})  = \frac{e^2}{h}\pi \Gamma\rho(0) $ \cite{Meir92} in terms of the dot spectral 
function $ \rho  (\omega)  = -\frac{1}{\pi} {\rm Im} \,{\mathcal{G}}(\omega+i0^+) $.
In order to implement the fRG, 
an infrared cutoff in the bare propagator  is introduced
$  {\mathcal{G}}_{0}^\Lambda(i\omega)
  ={\mathcal{G}}_{0}(i\omega)\Theta(| \omega |-\Lambda)$.
As the cutoff scale $\Lambda$ is gradually lowered, more and more
low-energy degrees of freedom are included, until finally the original
model is recovered for $\Lambda\to 0$.  Changing the cutoff scale
leads to an infinite hierarchy of flow equations for the vertex
functions.  
In the static approximation the flow equation for the effective level position 
$V^\Lambda = V_{\rm g}  + \Sigma^\Lambda$ reads
\begin{equation}
  \partial_\Lambda V^\Lambda  = \frac{U^\Lambda V^\Lambda/\pi}
  {(\Lambda+\frac{\Gamma}{2})^2+(V^\Lambda)^2}
\end{equation}
with the initial condition $V^{\Lambda=\infty} = V_{\rm g}$ \cite{KEM06}.  In first approximation the 
two-particle vertex $U^\Lambda\equiv
U^{\Lambda=\infty}=U$ equals the bare Coulomb interaction. At the end of the flow, the renormalised
potential $V = V^{\Lambda=0}$ determines the conductance 
$G(V_{\rm g}) = 2\, \frac{e^2}{h}\frac{\Gamma^2}{4V^2 + \Gamma^2}$.
Results for different values of $U/\Gamma$ are shown in figure~\ref{fig:fig3}, together with the 
occupation of the dot. 

For $\Gamma \ll U$ the resonance exhibits a plateau \cite{Gerland00}.
In this region the occupation is close to $1$ while it sharply
rises/drops to $2$/$0$ to the left/right of the plateau.  For
asymmetric barriers the resonance height is reduced to $2\,\frac{e^2}{h}\frac{4\Gamma_{\rm L}
\Gamma_{\rm R}}{(\Gamma_{\rm L}+\Gamma_{\rm R})^2}$ \cite{Hewson93,Gerland00}.
Focusing on strong couplings $U \gg \Gamma$, the solution of the above flow equation 
$  V \simeq V_{\rm g}\, e^{  -\frac{2U}{\pi \Gamma}}$ describes the
exponential pinning of the spectral weight at the chemical potential for small $| V_{\rm g} |$ and
the sharp crossover for a $V$ of order $U$. While already the first order in the flow-equation 
hierarchy captures the correct physical behavior, the inclusion of the renormalisation of the 
two-particle vertex improves the quantitative accuracy of the results. For details on the 
parametrization and extensions including dynamical properties we refer to Refs. 
\cite{Hedden04,KEM06,Karrash08,Jakobs10,Jakobs10a}. 
In presence of a magnetic field the Kondo resonance splits into two peaks with a dip in
$G(V_{\rm g})$ at $V_{\rm g}=0$, providing a definition of the Kondo scale as the magnetic field required to 
suppress the total conductance to one half of the unitary limit.  
For the single dot at $T=0$ the conductance and transmission phase are
related by a generalized Friedel sum
rule to the dot occupancy \cite{Hewson93} by 
$G=2\,\frac{e^2}{h}\sin^2{({\pi\over 2} \left< n\right> )}$ and 
$\alpha={\pi\over 2} \left< n \right>$.   
For gate voltages within the conductance
plateau the dot filling is $1$ and the phase is $\frac{\pi}{2}$. 
The description of transmission phases for more complex setups
will be discussed in the next subsection.

\subsection{Mesoscopic to universal crossover of the transmission phase of
  multi-level quantum dots}

\begin{figure}[t]
\center{\includegraphics[width=.6\textwidth,clip]{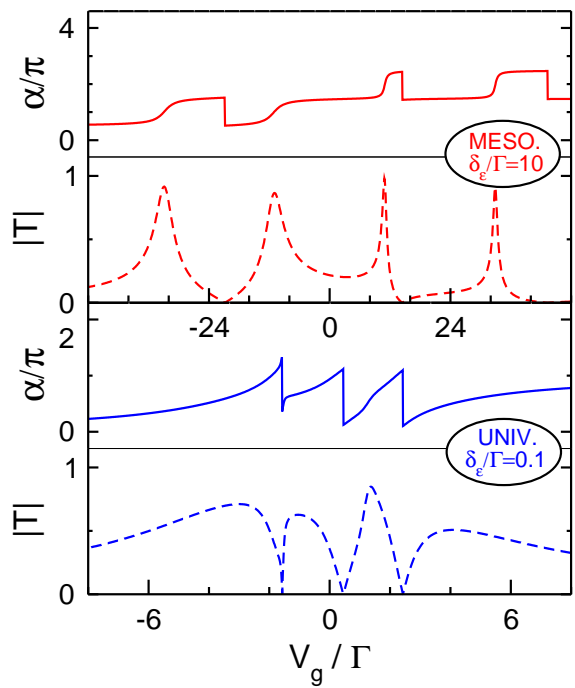}}
\caption[]{(Color online) Transmission amplitude $|T(V_{\rm g})|$ and phase $\alpha (V_{\rm g})$ for $U/\Gamma=1$ and $N=4$ equidistant levels with spacing $\delta_{\epsilon}$.
  Decreasing $\delta_{\epsilon}/\Gamma$ leads to a crossover from
  mesoscopic to universal behavior.
  \label{fig2}}
\end{figure}

The following application to a multi-level quantum dot illustrates the strength of the fRG approach, 
the flexibility and simple implementation, in the description of the
intriguing phase-lapse behavior observed in experiments by the group 
of Heiblum at the Weizman Institute \cite{Yacoby95,Schuster97}.
The transmission amplitude and phase $T = |T|e^{i \alpha}$
of electrons passing through a quantum dot embedded in an Aharonov-Bohm 
geometry showed a series of peaks in $|T|$ as a function of a plunger
gate voltage $V_{\rm g}$ shifting the dot's single-particle
energy levels. Across these Coulomb blockade peaks
$\alpha (V_{\rm g})$ continuously increased
by $\pi$, as expected for Breit-Wigner resonances.
In the valleys the transmission phase revealed ``universal'' jumps by $\pi$ for large dot occupation 
numbers. In contrast,  for small dot fillings the appearance 
of a phase lapse depends on the dot parameters in the ``mesoscopic" regime.
From the theoretical side, the observed behavior is captured by an fRG computation of the transport
through a multi-level quantum dot with local Coulomb interactions \cite{Karrasch07a,Karrasch07b}. 
An essential aspect in the description of the generic gate-voltage dependence of the transmission lies 
in the feasibility of a systematic analysis of the whole parameter space.
The interaction is taken into account approximately, but a comparison to numerically exact NRG data 
for special parameters proves the fRG results to be reliable as long as the 
interaction parameter and the number of almost degenerate levels do not become 
too large simultaneously. 
The results are shown in figure~\ref{fig2}, for spin-degenerate levels we refer to 
Ref. \cite{Karrasch07b}.  

Assuming level spacings of the order of the level width or smaller 
for large dot fillings (similar to a hydrogen atom with decreasing level spacing for increasing quantum number) 
and well separated levels for dots occupied by 
only a few electrons, the 
obtained results are consistent with the experimental findings in both regimes. In the ``mesoscopic'' 
regime the Coulomb blockade yields an increase in the separation of the transmission peaks. 
The behavior in the valleys depends on the details of the level-lead 
couplings, and can be continuous or discontinuous with a phase lapse of $\pi$ (see upper panels in 
figure~\ref{fig2}). For several almost degenerate levels in the ``universal'' regime the 
hybridization leads to a single broad and several narrow effective levels (Dicke effect 
\cite{Berkovits04}).  In presence of the Coulomb repulsion, the 
gate-voltage dependence of the broad level is significantly reduced, in
contrast to the narrow levels crossing the broad level close to the chemical potential. 
The combination of these effects induces well separated transmission peaks, absent without 
interaction. The ``universal'' phase lapse can be understood as a result of the Fano effect of the 
effective renormalised levels, where $\pi$ phase lapses appear in resonance phenomena
(see lower panels in figure~\ref{fig2}). 
The mechanism leading to the phase lapses reflects also in the dot level occupancies, the 
broad level being filled and emptied via the narrow levels \cite{Silvestrov00,Silvestrov01}. 
The detailed understanding of the charging of a narrow level coupled to a wide one in presence of 
interactions is of interest also in 
connection with experiments on charge sensing \cite{Johnson04}. From an fRG analysis emerges a
single-parameter scaling behavior of the characteristic energy scale for the charging of the narrow 
level with an interaction-dependent scaling function \cite{Karrasch09}. 

\subsection{Outlook}
The described fRG approach presents a reliable and promising tool for
the investigation of correlation and quantum fluctuation effects in quantum dots,
the computational effort being comparable to a mean-field calculation.
In contrast to the latter the fRG does not lead to unphysical artefacts. 
Besides the Kondo effect and transmission phases in 
multi-level dots, as explained above, 
also the competition between the Kondo effect and interference phenomena in more complex 
quantum-dot systems can be described \cite{KEM06, Meden06, sup1,sup2,sup3}. 
Furthermore, correlation effects for a large number of single-particle levels can be discussed up
to the limit of one-dimensional quantum wires. These applications involve
inhomogeneous Luttinger liquids \cite{MMSS02a,Andergassen04,EMABMS04,Andergassen06}, which will be
discussed in section \ref{sec:wires}, as well as the influence of Luttinger liquid leads coupled to 
a quantum dot \cite{AEM06}. 
Challenging extensions and theoretical developments of the fRG involve the analysis of inelastic processes 
to describe dynamical properties,
and the inclusion of higher-order contributions to access the physics at large $U/\Gamma$.
A fundamental question concerns the combination with the Keldysh formalism to describe nonequilibrium problems addressed in the next section.

\section{Quantum fluctuations in nonequilibrium and time evolution}
\label{sec:fluct_nonlinear}

In this section we discuss quantum fluctuations in quantum dots or molecules
in the presence of a finite bias voltage together with the time evolution
into the stationary state. As in the previous section the quantum fluctuations
are induced by the coupling to the reservoirs. If the maximum of temperature $T$ and 
the distance to resonances $\delta$ decreases, correlations effects from the Coulomb 
interaction lead to an increased renormalisation of the couplings and the excitation 
energies $h_i$ of the quantum dot (like e.g. magnetic field, level spacing, 
single-particle energy, etc.). Renormalisation group (RG) methods, 
which expand systematically in the reservoir-system
coupling (in contrast to the fRG method described in the previous section \ref{sec:fluct_linear}, 
where an expansion in the Coulomb interaction is performed), 
show that these renormalisations are typically 
logarithmic or power laws. In nonequilibrium or for the time evolution, the 
voltage $V$ and the inverse time scale $1/t$ are two new energy scales, which can 
cut off the RG flow. New physical phenomena emerge,
which we will illustrate in this section by considering very basic two-level
quantum dots, like two spin states (Kondo model) or two charge states 
(interacting resonant level model (IRLM)). Conceptually, one has to distinguish between
the two different regimes of strong and weak quantum fluctuations. In strong coupling,
an expansion in the renormalised coupling is no longer possible, and,
except for the case of strong charge fluctuations in the IRLM or for the case of 
moderate Coulomb interactions (which can be treated with fRG methods), the nonequilibrium case
remains a fundamental yet unsolved problem. Especially for the 
case of strong spin fluctuations, represented by the nonequilibrium Kondo model
or the nonequilibrium single-impurity Anderson model, an analytical or
numerical solution is still lacking. Even very basic questions, like e.g. the splitting of
the Kondo peak in the spectral density by a finite bias voltage are not yet
clarified. In weak coupling, where a controlled expansion in the 
renormalised coupling is possible, many results are
already known. The simplest ansatz is to take the bare
perturbation theory, as explained in section~\ref{sec:molecules}, use a Lorentzian
broadening of the energy conservation (induced by relaxation and
decoherence rates), and replace the bare couplings and excitation energies
by the renormalised ones obtained from standard equilibrium poor man scaling (PMS) 
RG equations cut off by the
maximum $\Lambda_{\mathrm{c}}=\max\{T,h_i,V,1/t\}$ of all physical energy scales.
However, it turns out that this approach is not sufficient.
In contrast to temperature, the bias voltage is not an infrared cutoff since it 
tunes the distance to resonances, where new physical processes, like e.g.
cotunneling (see section~\ref{sec:molecules}), can occur. Therefore,
the correct cutoff is not the voltage but the distance $\delta_i$ to resonances,
where RG enhanced contributions occur. Furthermore,
at resonance $\delta_i=0$, these contributions are cut off by relaxation
and decoherence rates $\Gamma_i$, i.e. one should use the energy scale
$|\delta_i+i\Gamma_i|$ as cutoff parameter for the couplings. An RG approach
has to be developed to reveal how the various cutoff parameters 
influence the couplings. It turns out that the rates $\Gamma_i$ are
{\it transport rates}, which have to be determined from a kinetic (or quantum
Boltzmann) equation, in contrast to rates describing the decay of a
local wave function into a continuum. Therefore, the RG has to be combined
with kinetic equations. The transport rates $\Gamma_i$ depend
also on voltage and are important parameters to prevent the system from 
approaching the strong coupling regime for voltages larger than the strong 
coupling scale $T_c$. Furthermore, it turns out
that terms can occur in the renormalised perturbation theory which are
not present in the bare one, like e.g. the renormalisation
of the magnetic field in linear order in the coupling for the
Kondo model (similar effects can also happen in higher orders for
generic models). For the time evolution, these terms are of particular
interest in the short-time limit, since in this regime they are cut off by the inverse 
time scale $1/t$ and lead to universal time evolution. At large times, 
non-Markovian parts of the dissipative kernel of the kinetic equation 
lead to many other interesting effects. Among them are
unexpected oscillation frequencies involving the voltage, unexpected
decay rates, power-law behavior, and a different cutoff behavior at 
resonances compared to stationary quantities. In particular, for metallic
reservoirs with a constant density of states, it can be shown that all 
local physical observables decay exponentially accompanied
possibly by power-law behavior. Since all these effects occur already
for very basic two-level systems, it has to be expected in the future 
that many more interesting effects will be found with respect to the physics of
quantum fluctuations in nonequilibrium.

From a technical point of view the description of correlated
quantum dots in the presence of quantum fluctuations in 
nonequilibrium is a very challenging problem and a playground for
the development of new analytical and numerical methods. 
Concerning numerical methods, numerical renormalisation group (NRG) 
with scattering waves \cite{Anders08}, 
time-dependent NRG (TD-NRG) \cite{Anders05,Anders06,Anders07,Roosen08},
time-dependent density matrix renormalisation group (TD-DMRG) 
\cite{Heidrich-Meisner09,Boulat08a, Heidrich-Meisner09,Daley04,White04,Schmitteckert04}, 
quantum Monte Carlo (QMC) with complex chemical potentials \cite{Han07}, 
QMC in nonequilibrium \cite{Schmidt08,Werner09a,Werner09b},
and iterative path-integral approaches (ISPI) \cite{Weiss08} have been developed.
However, the efficiency of these methods is often not satisfactory in the regime 
of either strong Coulomb interaction, large bias voltage, or long times. 
Exact analytic solutions exist for some special cases 
\cite{Lesage98,Schiller00,Komnik09,Boulat08a,Boulat08b} and
scattering Bethe-ansatz methods have been applied to the IRLM \cite{Mehta06}.
Other analytical methods are mainly based on RG approaches. Besides 
conventional PMS \cite{Glazman05}, improved frequency-dependent RG schemes 
\cite{Rosch01,Rosch03,Rosch05}, flow equations \cite{Kehrein05}, and 
fRG methods \cite{Jakobs03,Gezzi07,Jakobs07,Jakobs10,Jakobs10a,Karrasch10}
have been used. Whereas the latter expands systematically in the Coulomb interaction,
we will introduce in this section a formally exact RG method, which expands
systematically in the reservoir-dot coupling. It 
uses the perturbation theory described in section~\ref{sec:molecules} and
allows for a direct calculation of the kernel of the kinetic equation \eref{eq:kineq}
in Liouville space, the so-called real-time RG method (RTRG) \cite{Schoeller00a,Schoeller00b,Korb07},
which recently has been technically optimized and formulated in pure frequency 
space (RTRG-FS) \cite{Schoeller09a}. The method has been successfully 
applied to calculate analytically all static and dynamical aspects of weak
spin fluctuations for the anisotropic Kondo model at finite magnetic field 
\cite{Schoeller09b,Schuricht09,Pletyukhov10} and strong charge fluctuations for 
the IRLM \cite{Andergassen10}. Many aspects of these 
models can also be derived with other methods, which will be mentioned in the 
following at the appropriate places.

\subsection{Real-time RG in frequency space (RTRG-FS)}
\label{subsec5:rtrg_fs}

The aim of RTRG-FS is to combine the diagrammatic expansion in Liouville space,
as described in section~\ref{sec:molecules}, with RG to resum systematically
infrared divergences. We consider a time translational invariant system and
use as an initial condition at $t_0=0$ that the dot and the reservoirs are decoupled. The kinetic equation \eref{eq:kineq}
can be written in Laplace space as
\begin{equation}
\label{eq:kineq_laplace}
\tilde{p}(z)\,=\,{i\over z - L_{\mathrm{D}}^{\mathrm{eff}}(z)}\,p(t=0)\quad,
\end{equation}
where $\tilde{p}(z)=\int_0^\infty dt \,e^{izt}p(t)$ is the reduced density
matrix of the dot in Laplace space and $L_{\mathrm{D}}^{\mathrm{eff}}(z)=L_{\mathrm{D}} + \tilde{\Sigma}(z)$,
where $\tilde{\Sigma}(z)$ denotes the Laplace transform of the kernel
$\Sigma(t-t')=\Sigma(t,t')$. The effective dot Liouville operator $L_{\mathrm{D}}^{\mathrm{eff}}(z)$
contains all reservoir degrees of freedom and is of dissipative nature. 
The qualitative dynamics can be analyzed from the analytic structure of $\tilde{p}(z)$,
which is an analytic function in the upper half of the complex
plane with poles and branch cuts only in the lower half. The single poles
located at $z^i_{\mathrm{p}} = h_i - i\Gamma_i$ correspond to exponential decay with
oscillation frequencies $h_i$ and decay rates $\Gamma_i$. From the pole
at $z=0$, the stationary state can be obtained via the solution
of $L_{\mathrm{D}}^{\mathrm{eff}}(i0^+)p=0$. Due to non-Markovian terms arising from the
$z$-dependence of the Liouvillian, the weight of these poles is changed by 
$Z$-factors and branch cuts can occur. 
For a constant density of states in the leads (normal metallic case)
it can be shown generically \cite{Pletyukhov10,Andergassen10} that
the $Z$-factor leads to universal short-time behavior, 
whereas the branch cuts give rise to an exponential decay $\sim e^{-iz_{\mathrm{b}}^i t}$ 
accompanied by power-law behavior. $z_{\mathrm{b}}^i$ denotes the position of
the branching point, which is shifted from some pole position 
by multiples of the electrochemical potentials of the reservoirs leading to oscillation 
frequencies involving the voltage. From the point of view of error correction
schemes in quantum information processing it is quite important to understand
these corrections to Markovian behavior \cite{Peskill98,DiVincenzo05,Fischer09}.

It is unique to the RTRG-FS method that it provides direct access to the important
quantity $L_{\mathrm{D}}^{\mathrm{eff}}(z)$. By systematically integrating out the energy scales of the reservoirs
step by step, a formally exact RG equation can be derived for $L_{\mathrm{D}}^{\mathrm{eff}}(z)_\Lambda$, where
all reservoir energy scales beyond $\Lambda$ are included. This RG equation is coupled to
other RG equations for the couplings. Similar schemes can be developed for the
calculation of the transport current \cite{Schoeller09a} and correlation functions 
\cite{Schuricht09}. All RG equations involve resolvents similar to the one occurring in 
\eref{eq:kineq_laplace} where $z$ is replaced by $\Lambda$ together with other physical energy scales. 
As a consequence, it can be shown that, besides temperature, 
each term of the RG equation has a specific cutoff scale $\Lambda_i$, 
which is generically of the form
\begin{equation}
\label{eq:cutoff}
\Lambda_i\,=\,|E~+~\sum_j n_j\mu_{\alpha_j}~-~h_i~+~i\Gamma_i| 
~\equiv~ |\delta_i~+i\Gamma_i|\quad.
\end{equation}
Here, $E$ is the real part of the Laplace variable, $n_j$ are integer numbers, 
and $\mu_\alpha$ denotes the electrochemical potential of reservoir $\alpha$.
It shows that the cutoff scale is given by the distance $\delta_i$ to
resonances. Furthermore, it provides the generic proof that, at resonance $\delta_i=0$,
the cutoff scale is given by the corresponding rate $\Gamma_i$. This issue was
under debate for some time because it was speculated that electrons tunneling in
and out via the same reservoir correspond to low-energy processes, which could
possibly lead to a strong coupling fixed point even in the presence of a finite 
bias voltage \cite{Coleman01}. However, it was argued that voltage-induced decay 
rates prevent the system from approaching the strong coupling regime
\cite{Kaminski00,Parcollet02,Rosch01}. The 
microscopic inclusion of decay rates as cutoff scales into nonequilibrium RG 
methods was achieved within RTRG \cite{Schoeller00a,Schoeller00b,Korb07},
flow equation methods \cite{Kehrein05}, and RTRG-FS \cite{Schoeller09a}.

\begin{figure}
  \centering
  \includegraphics[width=160mm]{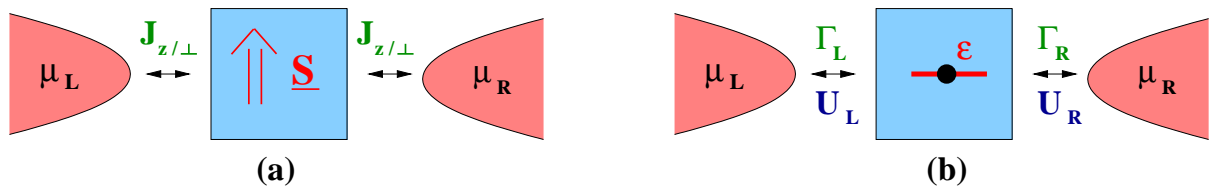}
  \caption{(color online)
    Two fundamental quantum dot models. (a) is the Kondo model,
    a spin-${1\over 2}$ coupled via exchange couplings $J_{z,\perp}$ to two reservoirs. (b) is 
    the IRLM, a spinless $1$-level quantum dot coupled via tunneling rates $\Gamma_{\mathrm{L,R}}$ and
    Coulomb couplings $U_{\mathrm{L,R}}$ to two reservoirs. The electrochemical potentials are
    given by $\mu_{\mathrm{L/R}}=\pm V/2$
}
\label{fig:kondo_irlm}
\end{figure}

\subsection{Applications}
\label{subsec5:applications}

The two models used to illustrate the basic physics of spin and charge fluctuations
are sketched in figure \ref{fig:kondo_irlm}. One model is the Kondo model 
at finite magnetic field $h$ already discussed in section~\ref{sec:fluct_linear}, 
where a spin-$1/2$ couples via anisotropic exchange couplings $J_{z/\perp}$ 
to the spins of two reservoirs. We have assumed a symmetric coupling to the leads and
note that during the exchange it is also allowed that a particle is transferred between
the reservoirs. The model results from the Coulomb blockade regime of a quantum 
dot with one level, where charge fluctuations are frozen out and only the spin can fluctuate. 
This leads to an effective band width of the reservoirs of the order of the charging energy $U$.
Anisotropic exchange couplings can be realized for a molecular magnet, see 
section~\ref{sec:molecules}. The other model is the IRLM, where the quantum dot consists of a 
single spinless energy level at position $\epsilon$. The dot interacts with the reservoirs 
via tunneling processes, which are characterized by the tunneling rates
$\Gamma_\alpha = 2\pi\rho_\alpha |T_\alpha|^2$, with $\alpha=\mathrm{L},\mathrm{R}$. In addition, there
is a Coulomb interaction $u_{\alpha}$ between the first site of the reservoir leads and 
the quantum dot, which are characterized by the dimensionless parameter 
$U_\alpha=\rho_\alpha u_\alpha$. In the following we denote the bare parameters by
a super-index ${}^{(0)}$.

The two models have in common that due to spin/charge conservation the effective 
Liouvillian has the same matrix structure. There are three nonzero poles of the 
resolvent \eref{eq:kineq_laplace} at $z^1_{\mathrm{p}}=-i\Gamma_1$ and $z^\pm_{\mathrm{p}}=\pm h-i\Gamma_2$.
For the Kondo model, $\Gamma_{1/2}$ corresponds to the spin 
relaxation/decoherence rate and $h$ is the renormalised magnetic field. For
the IRLM, $\Gamma_1$ is the charge relaxation rate, $\Gamma_2$ describes the broadening
of the local level, and $h\equiv\epsilon$ is the renormalised level position.
Denoting the two eigenstates of the dot by 
$\pm\equiv\uparrow,\downarrow\equiv 1,0$, the matrix 
elements $L_{\mathrm{D}}^{\mathrm{eff}}(i0^+)_{ss,s's'}=-iss'W^{-s'}$ of the Liouvillian involve the
rates $W^s$ for the process $-s\rightarrow s$. The stationary 
occupations are given by $p_s=W^s/W$, which contain already most of the interesting
nonequilibrium physics. Similar rates can be defined to calculate the current.
We consider temperature $T=0$ from now on to reveal the physics of quantum fluctuations.

\subsubsection{Kondo model.} 
\label{subsubsec5:kondo}

We consider first the stationary case in the weak coupling regime 
$\Lambda_{\mathrm{c}}=\max\{V,h\}\gg T_{\mathrm{K}}$,
where the Kondo temperature $T_{\mathrm{K}}\equiv T_c$ is the strong coupling scale
for spin fluctuations. The renormalised couplings
from PMS cut off at the scale $\Lambda_{\mathrm{c}}$ are denoted by $J_{z/\perp}$.
For the isotropic case, they are explicitly given by $J=1/(2\ln(\Lambda_{\mathrm{c}}/T_{\mathrm{K}}))$, with
$T_{\mathrm{K}}\sim D\sqrt{J^{(0)}}\,e^{-1/(2J^{(0)})}$ ($D\sim U$ denotes the effective
band width of the reservoirs).

In lowest order in $J_{z/\perp}$, one obtains a \lq\lq golden rule'' like expression for 
the rates \cite{Schoeller09b} 
\begin{equation}
\label{eq:rates_kondo}
W^s~=~{\pi\over 2}J_\perp^2\left\{4h\delta_{s-}+(V-h)\theta_{\Gamma_2}(V-h)\right\}\quad,
\end{equation}
where $h,V>0$ and $\theta_\Gamma(\omega)=1/2+(1/\pi)\arctan(\omega/\Gamma)$ is a step function
broadened by $\Gamma$. The latter corresponds to a Lorentzian broadening
of the energy conservation law by quantum fluctuations.
The spin relaxation/decoherence rates and the renormalised magnetic field are given by
% \begin{eqnarray}
% \label{eq:gamma_12_kondo}
% \fl \Gamma_1~=~{\pi\over 2}J_\perp^2\left\{4h~+~(V-h)\theta_{\Gamma_{12}}(V-h)\right\} \quad,\quad
% \Gamma_2~=~{1\over 2}\Gamma_1~+~{\pi\over 2}VJ_z^2 \quad,\\
% \label{eq:h_kondo}
% \fl h~=~(1-J_z+J_z^{(0)})h^{(0)}~-~hJ_\perp^2\ln({\Lambda_{\mathrm{c}}\over|h+i\Gamma_{12}|})~+~
% {1\over 2}(V-h)J_\perp^2\ln({\Lambda_{\mathrm{c}}\over|V-h+i\Gamma_{12}|}) \,,
% \end{eqnarray}
\numparts
\begin{eqnarray}
\label{eq:gamma_12_kondo}
\Gamma_1 &=& {\pi\over 2}J_\perp^2\left\{4h+(V-h)\theta_{\Gamma_{12}}(V-h)\right\}\quad,
\\
\Gamma_2 &=& {1\over 2}\Gamma_1+{\pi\over 2}VJ_z^2\quad,
\end{eqnarray}
\endnumparts
where $\Gamma_{12}=\Gamma_1-\Gamma_2$ and
\begin{eqnarray}
\label{eq:h_kondo}
h & = & \left(1-J_z+J_z^{(0)}\right)h^{(0)}
\nonumber
\\
  &   & 
- hJ_\perp^2\ln {\Lambda_{\mathrm{c}}\over|h+i\Gamma_{12}|}
+ {1\over 2}(V-h)J_\perp^2\ln {\Lambda_{\mathrm{c}}\over|V-h+i\Gamma_{12}|}\quad,
\end{eqnarray}
where unimportant terms $\sim O(J^2)$ without a logarithm have been
left out for $h$.

The expression for the rates $W^s$ up to $O(J^2)$ can be interpreted as follows.
At high energies $\Lambda>\Lambda_{\mathrm{c}}$, the voltage and magnetic field are not relevant 
and the exchange couplings are renormalised according to the PMS
equations. This leads to an effective band width $\Lambda_{\mathrm{c}}$ for the reservoirs
with renormalised exchange couplings cut off at $\Lambda_{\mathrm{c}}$. The precise value of
$\Lambda_{\mathrm{c}}$ is not important since the renormalisation is logarithmic. After this
step one uses lowest order perturbation theory with broadened energy
conservation leading to the result \eref{eq:rates_kondo} for the rates.
However, this interpretation does not work for the rates in $O(J^3)$ and it fails for 
the renormalised magnetic field $h$ already in $O(J)$ and $O(J^2)$.
We see that $h$ contains a term linear in $J_z$, 
which does not occur in perturbation theory, and was also discussed
e.g. in \cite{Garst05}. It shows that it is generically not possible to calculate 
coefficients of certain orders in the renormalised couplings by comparing with bare 
perturbation theory. In addition, we find logarithmic corrections in $O(J^2)$.
As was already discussed generically by \Eref{eq:cutoff}, they occur at resonances
$h=0$ or $V=h$. As was emphasized in \cite{Glazman05}, they are of $O(J^2\ln J)$
at resonance and remain a perturbative correction for $J\ll 1$. They occur because the
exchange couplings are not completely cut off by $\Lambda_{\mathrm{c}}$ but partially by
smaller cutoff scales. The calculation of the prefactor of the logarithmic terms is
quite subtle and follows from the structure of the RG equations, where each term
has its own cutoff scale $\Lambda_i$, see \Eref{eq:cutoff}. Similar 
logarithmic corrections occur in $O(J^3)$ for the rates
and for $\Gamma_{1/2}$. In turns out that the logarithmic terms for $h$ and $\Gamma_{1/2}$ 
contain an unexpected cutoff scale $\Gamma_{12}=\Gamma_1-\Gamma_2$. This is generic for all 
quantities entering the time evolution. It occurs because the pole
positions of the resolvent \eref{eq:kineq_laplace} have to be calculated self-consistently.

\begin{figure}
  \centering
  \includegraphics[width=80mm]{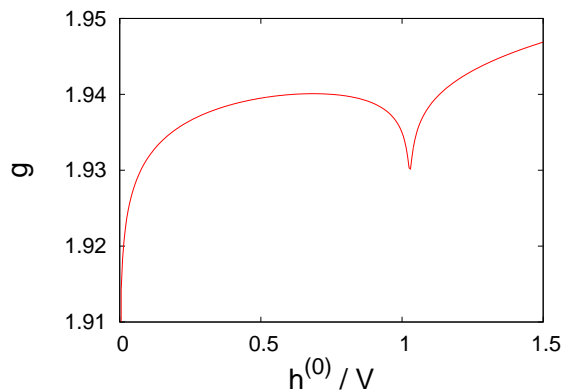}
  \caption{g-factor $g=2\,dh/d h^{(0)}$ for the isotropic Kondo model with
    $V=10^{-4}D$, $T_{\mathrm{K}}=10^{-8}D$.}
    \label{fig:g_factor}  
\end{figure}

Logarithmic enhancements have been studied for the magnetic susceptibility and for the
conductance also in \cite{Rosch03,Rosch05,Fritsch10}, using slave particle and flow equation 
methods. The logarithmic terms in the magnetic field have first been calculated in 
\cite{Schoeller09b} using RTRG-FS. They lead to a suppression of the renormalised 
g-factor as function of $h/V$ at $h=0$ and $h=V$, i.e. a {\it nonequilibrium} induced effect,
see figure \ref{fig:g_factor}.
Experimentally, it is proposed to be measured by using a three-lead setup, where the 
third lead is a weakly coupled probe lead \cite{Schoeller09b}. A similar interesting 
nonequilibrium effect occurs for the logarithmic enhancement of the magnetic susceptibility 
$\chi(h/V)$ at $h=0$, as first proposed in \cite{Rosch03}.
All spin-spin correlation functions were calculated analytically in \cite{Schuricht09}
using RTRG-FS. Here, the frequency variable of the correlation function enters 
\Eref{eq:cutoff} as well. At resonances, kink structures and logarithmic terms were found 
for the imaginary and real part of response functions, respectively.

In \cite{Pletyukhov10}, the time evolution has been calculated from the inverse Laplace
transform of \Eref{eq:kineq_laplace}. In the short-time limit $t\ll\Lambda_{\mathrm{c}}^{-1}$,
the PMS equations are cut off by the energy scale $1/t$ leading to time dependent
exchange couplings $J_{z/\perp}^t$. In $O(J)$ only the
$Z$-factor from the linear $z$-dependence of $L_{\mathrm{D}}^{\mathrm{eff}}(z)$ is important. 
In terms of the Pauli matrices $\boldsymbol{\sigma}$, the local density matrix 
can be written as $p(t)={1\over 2}+\langle\bi{S}\rangle(t)\,\boldsymbol{\sigma}$ with
\begin{eqnarray}
  \label{eq:short_t}
\langle\bi{S}\rangle(t)  = Z_t~\langle\bi{S}\rangle(0) 
                                =  \left\{1-2(J_z^t-J_z^{(0)})\right\}\langle\bi{S}\rangle(0)
\quad.
\end{eqnarray}
Inserting the solution of the PMS equations, one obtains universal
logarithmic (power law) time evolution for the isotropic (anisotropic) case. A similar
result has been found for the longitudinal spin dynamics in \cite{Hackl09a,Hackl09b} for 
the special case of the ferromagnetic Kondo model, which was also confirmed by TD-NRG 
calculations \cite{Roosen08}. In \cite{Pletyukhov10} it was also shown that the short-time 
behavior of the conductance can be calculated from the golden rule expression
and replacing the exchange couplings by $J_{z/\perp}^t$.

\begin{figure}
  \centering
  \includegraphics[width=120mm,clip=true]{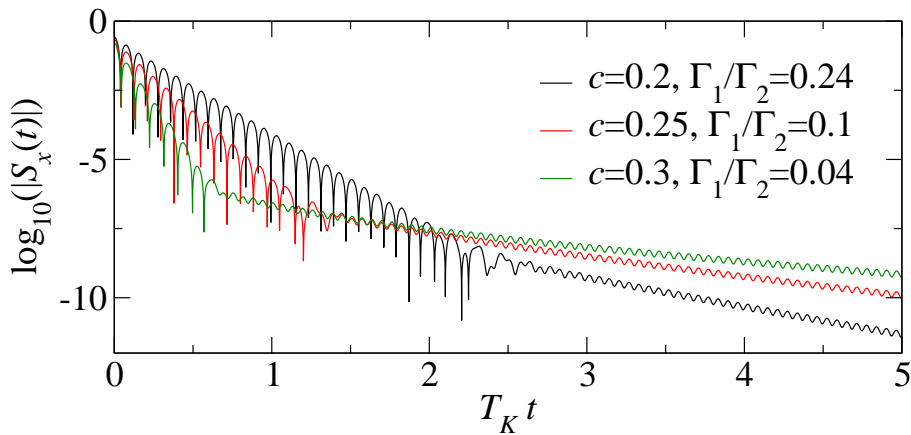}
  \caption{(Color online) $|S_x(t)|\equiv\langle|S_x(t)|\rangle$ in the anisotropic Kondo model for 
    $V=2h^{(0)}=100\,T_{\mathrm{K}}$ and various values of the anisotropy
    $c^2=(J_z^{(0)})^2-(J_\perp^{(0)})^2$, with $S_x(0)=1/2$, $S_y(0)=0$.  The
    dips have their origin in the oscillations of $S_x(t)$.}
  \label{fig:t_evolution}
\end{figure}

In the long-time limit $t\gg\Lambda_{\mathrm{c}}^{-1}$, the cutoff scale for the PMS couplings is 
given by $\Lambda_{\mathrm{c}}$. In leading (Markovian) order one obtains the usual exponential
behavior from the single poles of the resolvent \eref{eq:kineq_laplace}, where  
the longitudinal/transverse spin decays with $\Gamma_{1/2}$ and the transverse spin 
oscillates with $h$. Interesting non-Markovian corrections occur in $O(J^2)$, 
where logarithmic contributions $\sim J^2(z-z_{\mathrm{b}}^i)\ln(\Lambda_{\mathrm{c}}/(z-z_{\mathrm{b}}^i))$ 
in $L_{\mathrm{D}}^{\mathrm{eff}}(z)$ lead to branch cuts. The branching point of the logarithm is 
generically given by $z_{\mathrm{b}}^i=z_{\mathrm{p}}^j+nV$, i.e. is shifted by multiples
of the voltage from some pole position. As a consequence, an exponential behavior 
$\sim J^2 e^{-iz_{\mathrm{p}}^j t}e^{-inV}$ is obtained, accompanied by power-law behavior $\sim 1/t^2$ from the 
branch cut integral. The result explains why the voltage occurs generically in the oscillation frequency, which is
consistent with exact solutions at special Thoulouse points of two-level systems \cite{Komnik09}.
Furthermore, it shows that all terms decay exponentially but with unexpected decay rates and 
oscillation frequencies relative to the Markovian terms. This is illustrated in 
figure \ref{fig:t_evolution} by the time evolution of the transverse spin for 
the strongly anisotropic case $\Gamma_2\gg\Gamma_1$, which is typical for molecular magnets.
For short times, one obtains the expected result of oscillations with the magnetic field and
decay with the spin decoherence rate $\Gamma_2$. These terms decay quickly for large decoherence
rate. For longer times, a crossover is obtained
to an oscillation with the voltage $V$ and a decay with the opposite spin relaxation
rate $\Gamma_1$. Finally, we mention that close to resonances, i.e. for $|\delta|\ll\Lambda_{\mathrm{c}},1/t$, 
with $|\delta|=|V-h|,h$, logarithmic terms $\sim J^2\delta t\ln|(\delta+i\Gamma_{12})t|$ occur
for the time evolution of the transverse spin. Similar to the logarithmic terms in $h$ and
$\Gamma_{1/2}$, they are cut off by $\Gamma_{12}=\Gamma_1-\Gamma_2$.

\subsubsection{Interacting resonant level model}
\label{subsubsec5:irlm}

For the IRLM the PMS equations lead to a renormalised tunneling rate, given by the power-law
$\Gamma_\alpha=\Gamma_\alpha^{(0)}(D/\Lambda)^{g_\alpha}$, with exponent 
$g_\alpha=2U_\alpha-\sum_\beta U_\beta^2$. Cutting off this scale at 
$T_{\mathrm{c}}=\Gamma_L+\Gamma_R$, one defines self-consistently the strong coupling
scale $T_c$ for the importance of charge fluctuations. The level position
$\epsilon\equiv\epsilon^{(0)}$ and the Coulomb interaction $U_\alpha\equiv U_\alpha^{(0)}$
remain unrenormalised \cite{Borda07}. In lowest order in $\Gamma_\alpha$, one obtains
a golden-rule like expression for the rates 
\begin{equation}
\label{eq:rates_irlm}
W^s~=~\sum_\alpha W^s_\alpha \quad,\quad 
W^s_\alpha~=~\Gamma_\alpha \, \theta_{\Gamma_2}(s(\mu_\alpha-\epsilon))\quad,
\end{equation}
where $\mu_{\mathrm{L/R}}=\pm V/2$ are the electrochemical potentials of the leads.
The energy broadening is given by $\Gamma_2=\Gamma_1/2$ with the charge relaxation rate
$\Gamma_1=\sum_\alpha\Gamma_\alpha$. Since a single tunneling process changes the
charge, there is a unique cutoff parameter $\Lambda_{\mathrm{c}}^\alpha=|\mu_\alpha-\epsilon+i\Gamma_2|$
for the PMS tunneling rates. This cutoff scale will be used in the following to define $\Gamma_\alpha$.
For $|\mu_\alpha-\epsilon|\sim T_c$, we are in the regime of strong charge fluctuations.
We note the essential difference to the Kondo model, where several 
cutoff parameters are relevant. The current in lead $\alpha$ can be calculated from the
rate equation $I_\alpha={e^2\over h}2\pi(W^+_\alpha p_0-W^-_\alpha p_1)$ with the occupation 
probabilities $p_1=W^+/\Gamma_1$ and $p_0=W^-/\Gamma_1$. These results are quite 
specific to the IRLM and are due to its elementary form. They have been conjectured in \cite{Borda07} 
and later were confirmed microscopically by RTRG-FS \cite{Andergassen10} and fRG 
\cite{Karrasch10}. It is surprising that these results are consistent with NRG \cite{Borda07}
and TD-DMRG \cite{Boulat08a} calculations even for values of $U_\alpha~\sim O(1)$, and
capture all features obtained by field theoretical \cite{Boulat08a,Boulat08b}
and scattering Bethe ansatz \cite{Mehta06} approaches. In particular, negative differential
conductance is obtained for large voltages due to the power-law suppression of the
tunneling rates.

Defining $\Lambda_{\mathrm{c}}=\max\{\Lambda_{\mathrm{c}}^{\mathrm{L}},
\Lambda_{\mathrm{c}}^{\mathrm{R}}\}$ and expanding in $U_\alpha$, one finds
logarithmic enhancements close to the resonances $\Lambda_{\mathrm{c}}^\alpha=0$
\begin{equation}
\label{eq:log_irlm}
\Gamma_\alpha~=~\Gamma_\alpha^{(0)}\left({D\over\Lambda_{\mathrm{c}}}\right)^{g_\alpha}
\left(1~+~2U_\alpha\ln{\Lambda_{\mathrm{c}}\over|\mu_\alpha-\epsilon+i\Gamma_2|}~+~O(U^2)\right)\quad.
\end{equation}
It is important to notice that if the level is in
resonance with one of the reservoirs it is not with the other, i.e. exactly at resonance the
cutoff scale is $\Gamma_2$ for one rate and the voltage $V$ for the other rate. This has to
be contrasted to speculations that both rates are cut off by the geometric average
$\sqrt{\Gamma_2 V}$ \cite{Doyon07}, which leads to incorrect power-law exponent for the
on-resonance current as function of the voltage. Therefore, a microscopic determination
of the precise cutoff scales from nonequilibrium RG methods is very essential. 

The time evolution into the stationary state has been calculated for the IRLM in
\cite{Andergassen10}. The basic features already found for the Kondo model were
recovered, showing that these effects even hold for models with charge fluctuations. 
It turns out that the oscillation frequencies are generically given by
$h_i + \sum_j n_j\mu_{\alpha_j}$, with integer numbers $n_j$, i.e. the excitation 
energies $h_i$ of the dot are shifted by arbitrary multiples of the chemical potentials of the 
reservoirs. For the IRLM it turns out that the algebraic part
of the time-evolution is $\sim (1/t)^{1-g_\alpha}$, i.e. the exponent depends
on the Coulomb interaction strength.

\subsection{Outlook}
\label{subsec5:outlook}

The status of the field of quantum fluctuations in nonequilibrium for strongly
correlated quantum dots is that powerful techniques have been developed to study the
weak-coupling limit in a controlled way. Elementary models of spin and charge fluctuations
are well understood in this limit and the methods can now be applied to more complex
quantum dot models, such as discussed in section~\ref{sec:complex}.
However, two fundamental issues are still open.
First, the case of strong spin fluctuations, represented in its most elementary form
by the isotropic Kondo model in the limit $\mbox{max}\{T,h,V\}\sim T_{\mathrm{K}}$,
is one of the most fundamental unsolved problems. The case of strong
charge fluctuations at resonances has so far only been understood for the IRLM, where
it seems that the broadening of the level together with a PMS renormalisation
of the tunneling couplings captures the essential physics. Whether this holds also for more
complicated models has to be tested in the future. Secondly, most of the methods
used to describe nonequilibrium properties of quantum dots are parametrized by the
many-body eigenstates of the isolated quantum dot. Therefore, they can not be extended
easily to larger systems like multi-level quantum dots or quantum wires. Two exceptions
are the flow-equation method \cite{Kehrein05} and nonequilibrium fRG methods.
\cite{Jakobs03,Gezzi07,Jakobs07,Jakobs10,Jakobs10a,Karrasch10}. In fRG, as already 
described in section \ref{sec:fluct_linear},
a perturbative expansion in the Coulomb interaction is used and the vertices are parametrized
in terms of single-particle levels. Therefore, although fRG is restricted to the regime
of moderate Coulomb interactions, the potential of the method lies particularly
in the possibility to treat multi-level quantum dots and quantum wires. Preliminary studies have
tested the fRG for the IRLM and the single-impurity Anderson model (SIAM). For the IRLM, a 
static version was used, where 
the frequency dependence of the Coulomb vertex was neglected \cite{Karrasch10,Andergassen10}. 
An agreement was found with RTRG-FS and TD-DMRG \cite{Boulat08a}. 
Recently, a dynamic scheme with frequency-dependent 2-particle vertices has been developed
and was used to analyze the nonequilibrium SIAM in the regime of strong spin fluctuations
\cite{Jakobs10,Jakobs10a}. A good agreement with TD-DMRG, ISPI and QMC 
results was obtained for moderate Coulomb interactions \cite{Eckel10}. As will be described in the
next section, the static version of nonequilibrium fRG has also been applied to
quantum wires and it is a challenge for future research to generalize the dynamic
fRG scheme to quantum wires as well.

\section{Correlation effects in quantum wires}
\label{sec:wires}
\subsection{An introduction to Luttinger liquid physics}

A different type of correlation physics than discussed so far is found in (quasi) one-dimensional (1d) 
quantum wires at low temperatures. By this we mean electron systems confined in two spatial
directions such that only the lowest 1d subband is occupied. 
Compared to the quantum dots considered above they contain {\it many} correlated degrees of freedom and 
the single-particle level spacing of the wire becomes the smallest energy scale 
(``almost'' continuous spectrum; see below). The dominant 
quantum fluctuations are now driven by the interaction $U$ itself.
In fact, below we will first discuss isolated, translationally 
invariant wires. Still, as discussed in the main part of this section, 
the physics becomes even more interesting if the coupling to leads (or the 
coupling between two wires) is included.
Quantum wires are realized in 
single-walled carbon nanotubes \cite{Bockrath99,Yao99}, specifically designed semiconductor 
heterostructures \cite{Auslaender02,Auslaender05}, and in atom chains which form on 
certain surfaces \cite{Schaefer08}. A different class of quasi 1d systems  
are highly anisotropic (chain-like) bulk materials (for a recent review see e.g. \cite{Claessen07}).
Compared to the first type of systems 
it is less clear how a single wire of this class can be incorporated as an element in an electronic 
transport nanodevice and they will thus play a minor role in the present discussion.  

Although being far from experimentally realizable, translationally invariant models of 1d electrons with 
sizeable two-particle interaction 
(Coulomb repulsion) were already studied in the fifties and sixties. Exactly solvable 
models were constructed leading to a rather quick (compared to higher-dimensional correlated systems) 
gain in the understanding of correlation effects. Tomonaga showed that {\it all} excitations of 
a 1d correlated electron system at (asymptotically) 
low energies are {\it collective} in nature  (``plasmons'') rather than single-particle-like as in 
the three dimensional counterpart (Landau quasi-particles) \cite{Tomonaga50}. In particular, collective 
spin and charge density excitations travel with different speed. Quite often this difference in velocities 
of collective excitations is mistaken as being characteristic for 1d systems although it might also 
occur in higher dimensional {\it Fermi liquids}. Only when adding, that {\it no} additional 
quasi-particle excitations are possible {\it spin-charge separation} becomes a unique feature 
of 1d correlated electrons. As a second (related) interaction 
effect certain correlation functions display power-law behavior as a function of energy $\omega$ and 
momentum $k$ as first shown for the single-particle momentum distribution function $n(k)$ by Luttinger 
\cite{Luttinger63} and Mattis and Lieb \cite{Mattis65}. Using the concepts of statistical physics one 
can view 1d chains of electrons as being (quantum) critical, although the exponents 
depend on the details of the underlying single-particle model (band structure, band filling) as well as on the 
interaction strength. S\'{o}lyom \cite{Solyom79} and 
Haldane \cite{Haldane81} showed using {\it renormalisation group (RG) arguments} that the physics of the two 
exactly solvable models---being quite similar they are nowadays considered as a single model, the 
{\it Tomonaga-Luttinger model}---is generic for 1d interacting electrons as long as 
correlations do not drive the system out of the metallic phase (e.g. into a Mott insulator phase). This
led to the term {\it Luttinger liquid} (LL) physics for the above correlation effects. 
In particular, Haldane argued that all the exponents appearing 
in correlation functions of a {\it spin-rotational invariant} system with repulsive two-particle interaction
can be expressed in terms of a single model parameter dependent number $K < 1$, with $K=1$ for 
noninteracting electrons. Together with the velocities of the charge- and spin-density 
excitations $v_{\rm c}$ and $v_{\rm s}$, $K$ completely determines the low-energy physics. For a given microscopic 
model the strategy to obtain the low-energy physics is thus the following: one has to determine the three 
{\it Luttinger liquid parameters} as functions of the model parameters (ways to achieve this are e.g. 
discussed in \cite{Schoenhammer05}) and plug them into the analytic 
expressions for correlation functions computed within the Tomonaga-Luttinger model.   

Despite the intense effort (see e.g. \cite{Bockrath99,Yao99,Auslaender02,Auslaender05,Schaefer08,Claessen07}) a commonly 
accepted experiment which shows LL behavior beyond any doubts is still pending. This 
has to be contrasted to the situation in quantum dots discussed earlier where the appearance of 
Kondo physics as a consequence of local correlations has convincingly been shown (see above). The lack 
of clear cut experiments is partly related to the status of the theory. LL theory only 
makes predictions for the {\it asymptotic} low-energy behavior but does not provide the scales on 
which this sets in. In the construction of the Tomonaga-Luttinger model all microscopic scales as e.g. 
given by the band curvature \cite{Imambekov09} and the shape of the two-particle interaction \cite{Meden99} 
are disregarded. One can imagine that the upper energy scale 
$T_{\rm c}$ beyond which LL physics can be found becomes so small that power 
laws are masked by other scales such as the 
single-particle level spacing set by the length of the wire. 
A profound comparison to experiments requires a knowledge of $T_{\rm c}$ 
and possible lower bounds for LL physics
while methods which 
allow to extract these scale for microscopic models are rare. It is thus mandatory to develop new methods 
which capture LL physics and can directly be applied to microscopic models. Further down, we will 
return to this issue. We note in passing that the standard ab initio method---density functional 
theory---cannot be applied as the existing approximation schemes fail when it comes to LL physics. 
       
When using a quantum wire as part of a low-temperature nanodevice one has to be alert that 
correlation effects might alter the electronic 
properties. To understand this let us first imagine an electron tunneling from the tip of a scanning tunneling 
microscope into the bulk of a 1d quantum wire. Using Golden Rule-like arguments it becomes clear that 
the differential conductance $dI/dV$ computed from the tunneling current $I$ is determined by the product 
(more precisely the convolution) of the single-particle spectral function $\rho(\omega)$ (the ``local 
density of states'') of the wire and the tip. As the latter is constant at small energies, 
$dI/dV$ displays the LL power law of the wire spectral function 
$\rho(\omega) \sim |\omega-\mu|^{\alpha_{\rm bulk}}$, with the wires chemical potential 
$\mu$ and $\alpha_{\rm bulk} =(K+K^{-1}-2)/4$ \cite{Schoenhammer05}. The exponent changes to 
$\alpha_{\rm end} = (K^{-1}-1)/2$ if the electron tunnels into the end of a LL as it is e.g. the case in 
end-contacted wires \cite{Schoenhammer05}. Evidently, tunneling in and out of either the bulk or the end of a wire 
plays a crucial role when using it in a nanodevice.  

Secondly, we imagine a local inhomogeneity in the quantum wire. It might either appear 
(in a difficult to control way) in the process of producing the wire or 
might intentionally (in a controlled way) be introduced to design a functional device. As examples for the 
latter one can think of double barriers \cite{Postma01} defining a dot region with energy levels 
that can be tuned via a backgate (see the discussion on transport through quantum dots) or junctions 
of several quantum wires \cite{Fuhrer00,Terrones02}. One can easily imagine that because of the collective nature 
of the excitations in a LL any inhomogeneity with a (single-particle) $2 k_{\rm F}$-backscattering component, with 
$k_{\rm F}$ being the Fermi momentum, strongly affects the low-energy physics. A first step to substantiate 
this expectation is to compute the $2 k_{\rm F}$-component of the static density-density-response function 
$\chi(q)$ of the Tomonaga-Luttinger model \cite{Luther74}. For a noninteracting 1d electron system it diverges 
logarithmically when $q$ approaches $2 k_{\rm F}$ (Lindhard function). The divergence is enhanced and becomes power-law like
$\chi(q) \sim |q-2 k_{\rm F}|^{K-1}$ if the interaction is turned on. This shows that the inhomogeneity strongly 
couples to the system and linear response theory breaks down. This physics was further investigated applying
different RG methods to different models \cite{Kane92,Yue94,Nazarov03,Polyakov03,Enss05}. These studies show that even a single 
impurity at low energy scales effectively acts as if the chain is cut at the position of the impurity with 
open boundary conditions at the two end points. E.g. for temperatures $T \to 0$ the linear conductance 
$G(T)$ across the impurity is suppressed in a power-law fashion $G(T) \sim T^{2 \alpha_{\rm end}}$ as can again 
be understood using Fermis Golden Rule and the power-law scaling of the local spectral function at the end
of a quantum wire (tunneling from end to end of two LLs). Remarkably, the exponent is {\it independent} of the bare 
scattering potential. Also a Breit-Wigner resonance of $G$ as a function of 
the level position (modified by) $V_{\rm g}$ showing up in the double-barrier geometry 
is strongly altered by the interaction. In case of a ``perfect'' resonance with peak conductance $e^2/h$ 
(per spin), that is for equal left and right barriers (which is  experimentally difficult to realize), 
the resonance width becomes zero, while the resonance completely disappears, that is $G \to 0$ for {\it all} 
$V_{\rm g}$, in all other cases.      
 
The discussed physics of inhomogeneous LLs can be understood in terms of an effective single 
particle problem. Using RG techniques one can show that the interplay of a local inhomogeneity 
and two-particle correlations leads to 
an effective oscillatory and slowly decaying ($\sim 1/x$, with the distance $x$ from the inhomogeneity) 
single-particle scattering potential of range $1/\delta$, where $\delta$ is the 
{\it largest} of the relevant energy scales (e.g. temperature). 
The wave-length of the potential is set by the (common) chemical 
potential $\mu$ of the leads. Scattering off this (so-called 
Wigner-von Neumann potential) leads to the discussed power laws in transport with the exponent 
given by the amplitude of the potential.\cite{Yue94,Nazarov03,Polyakov03,Enss05}

In the remaining part of this section on coherent transport through quantum wires we describe four 
examples of recent attempts (by three of the present authors together with varying colleagues) to gain a 
detailed understanding of the interplay of correlations and local inhomogeneities in systems with 
different geometries. We believe that our insights should be kept in mind when designing transport 
setups to search for LL physics. They might also be of relevance when quantum wires (in the above defined 
sense) are used as elements in future nanoelectronic devices. In this case correlation effects might either 
be used to enhance the functionality or must be tuned away if they corrupt the latter. 

In all the examples  
we use the  {\it functional renormalisation group} (fRG) method \cite{Salmhofer01} 
(for a brief review see also  \cite{Meden08}) 
to (approximately) treat the Coulomb 
interaction. It was already introduced in section \ref{sec:fluct_linear}. Compared to other methods it has the advantages 
that it (i) can directly be applied to microscopic models (continuum or lattice; flexible modeling), (ii) 
captures all energy scales (not restricted to the asymptotic low-energy regime), and 
(iii) can be applied to systems with many correlated degrees of freedom. In the present context (iii) 
means  chains of realistic length (in the micrometer range) coupled to effectively 
noninteracting (Fermi liquid) leads. The aspects (i) and (ii) together assure that 
$T_{\rm c}$ and additional energy scales affecting LL physics can be investigated. 
As mentioned in section 
\ref{sec:fluct_linear}, the approximations inherent to the fRG approach 
are justified up to moderate renormalised two-particle interactions. In the implementation (static self-energy) 
used for quantum wires (i) the LL exponents come out correctly (only) to leading order in the two-particle 
interaction and (ii) {\it inelastic processes} generated by the two-particle 
interaction are neglected.  We will return to the latter in the outlook.

\subsection{The role of two-particle backscattering} 
 
\begin{figure}[tb]
  \centering
  \includegraphics[width=85mm,clip]{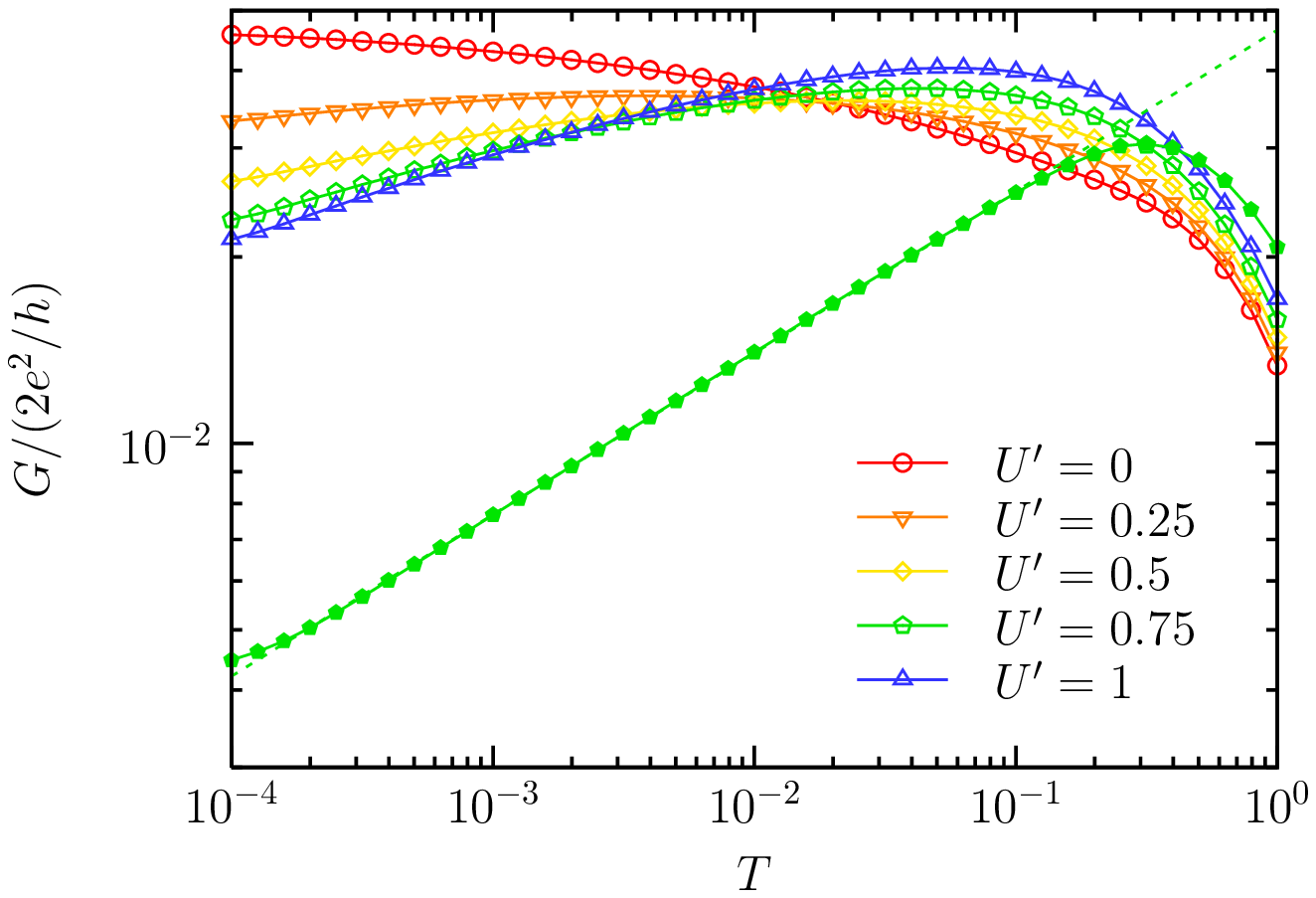}
  \caption{(Color online) Temperature dependence of the linear 
 conductance for the extended Hubbard model with $10^4$ sites 
 and a single site impurity (``$\delta$-impurity'') of strength 
 $V = 10$, for a Hubbard interaction $U = 1$ and various choices 
 of $U'$; 
 the density is $n=1/2$, except for the lowest curve,
 which has been obtained for $n=3/4$ and $U' = 0.65$ (leading to a 
 very small backscattering interaction);
 the dashed line is a power law fit for the latter parameter set. }
  \label{wires:fig1}
\end{figure}

In 1d systems at low energies and incommensurate fillings only {\it forward} (momentum 
transfer $q\approx0$) and {\it backward} (momentum transfer $q \approx 2 k_{\rm F}$) two-particle scattering 
is active. All other processes are suppressed due to momentum conservation and phase space 
restrictions (Fermi surface consists of only two points). The relative importance of the two processes 
is determined by the (real-space) range of the two-particle interaction, which in turn is set by the 
screening properties of the ``environment'', e.g. the substrate on which the carbon nanotube is placed.
If screening is strong, the two-particle interaction becomes short ranged in real space and 
backscattering is sizable and vice versa.   
Backscattering processes involving a spin up and a  spin down electron are not part of the 
Tomonaga-Luttinger model. For a translationally invariant LL it was shown using an RG analysis 
\cite{Solyom79} that these processes do not modify the asymptotic low-energy physics (they are 
``RG irrelevant''), but that they become irrelevant only on a scale which is {\it exponentially 
small} in the bare backscattering strength (they flow to zero only logarithmically). From this 
one can expect that the LL power laws only occur on exponentially 
small scales $T_{\rm c}$ also for inhomogeneous wires. 
For the {\it Hubbard model} with open boundaries it was shown that 
$T_{\rm c} \sim \exp{(-\pi v_{\rm F}/U)}$, where $U$ is the local 
Coulomb repulsion and  $v_{\rm F}$ the Fermi velocity \cite{Meden00}. 
In figure~\ref{wires:fig1} this behavior is exemplified for the $T$-dependence 
of the linear conductance $G$ of a wire with a single impurity 
described by the {\it extended Hubbard model} \cite{Andergassen06}. 
This lattice model consists of a standard tight-binding chain with nearest-neighbor hopping $t$, a local 
two-particle interaction $U$ as well as a nearest-neighbor one $U'$. 
Throughout the rest of this section we use $t$ as the unit of energy (that is we set $t=1$).
For a fixed band filling the relative strength of the 
forward and backward scattering can be modified by varying the ratio $U'/U$.               
The considered size of $N=10^4$ lattice sites corresponds to wires in 
the micrometer range, which is the typical size of quantum wires 
available for transport experiments. 
For $U'=0$ (strong backscattering) due to logarithmic corrections the conductance \emph{increases} as a function of 
decreasing $T$ down to the lowest temperatures in the plot, $T_{\rm c}$ being smaller than the latter. 
For increasing nearest-neighbor interactions $U'$ the (relative) importance 
of backscattering decreases and a suppression
of $G(T)$ at low $T$ becomes visible, but in all the data obtained
at quarter-filling $n=1/2$ the suppression is much less pronounced than 
what one expects from the asymptotic power law with exponent 
$2\alpha_{\rm end}$. 
By contrast, the suppression is much stronger and follows the
expected power law more closely if parameters are chosen such
that two-particle backscattering becomes negligible at low $T$,
as can be seen from the conductance curve for $n=3/4$ and 
$U'=0.65$ in figure~\ref{wires:fig1}. The value of $K$
for these parameters almost coincides with the one for another
parameter set in the plot, $n=1/2$ and $U'=0.75$, but the
behavior of $G(T)$ is completely different. To avoid interference 
effects discussed next, in the present setup the two leads are coupled 
``adiabatically'' to the wire, such that in the absence of the impurity 
the conductance would be $2e^2/h$ for all temperatures smaller than the band width \cite{Enss05}.

One can conclude that for 
wires in the micrometer range clear power laws can only be observed 
if screening is weak, leading to small two-particle 
backscattering \cite{Andergassen06,Yue94,Nazarov03,Polyakov03}.
Note that at $T \sim \pi v_{\rm F}/N$ 
finite size effects set in (the level structure becomes apparent), as can be 
seen at the low $T$ end of some of the curves in the figure.

\subsection{Overhanging parts}

In most transport experiments the wires are {\it not} end-contacted and the leads 
do {\it not} terminate at the contacts. This has to be contrasted to the modeling in which 
almost exclusively end-to-end contacted wires are considered. It is thus crucial to understand 
how overhanging parts of the wire and the leads alter the transport characteristics. 
Before tackling this problem we have to discuss transport through an interacting wire
with {\it two} contacts---above we only mentioned the single contact case with 
power-law scaling $G(T) \sim T^{\alpha_{\rm end}}$. One might be tempted to argue, that 
two contacts can be understood as two resistors in series, which then have to be added 
leading to $G(T) \sim T^{\alpha_{\rm end}}$ also for a (clean) wire with two contacts
to semi-infinite leads. 
Although the result is correct, as can be shown using (coherent) scattering 
theory \cite{Enss05,Jakobs07a}, this simple argument ignores, that adding of the resistance (of 
resistors in  series) only holds 
if the transport is {\it incoherent}, which is not the case at low 
temperatures as considered here. 

\begin{figure}[t]
  \centering
  \includegraphics[width=85mm,clip]{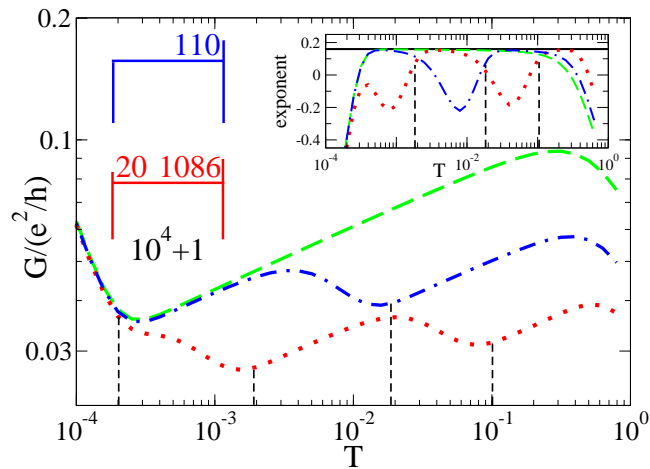}
  \caption{(Color online) Main plot: Linear conductance $G$ of an interacting wire with 
$U'=0.5$ of length $N=10^4+1$ as function of the temperature $T$. The coupling to the leads is 
located at the end of the wire and has 
a small amplitude $T_{\rm R}=T_{\rm L}=0.25$. Dashed line: Overhanging parts $N_{\rm lead}^{\rm L/R}=0$, that is end-to-end contacted 
wire. Dashed-dotted line: $N_{\rm lead}^{\rm L}=0, N_{\rm lead}^{\rm R}=110$.
Dotted line: $N_{\rm lead}^{\rm L}=20, 
N_{\rm lead}^{\rm R}=1086$. The vertical dashed lines terminating at the different curves 
indicate the corresponding crossover scales.  
At the lowest scale $T \sim 10^{-4}$ the power law is cut off by the level spacing $\sim 1/N$ 
of the disconnected wire.
Left inset: Setups studied (excluding the end-to-end contacted one).
Right inset: Effective exponents obtained from taking the logarithmic derivative of $G(T)$. 
The solid horizontal line indicates $\alpha_{\rm end}$.}
  \label{wires:fig2}
\end{figure}

We here exemplify the role of overhanging parts by considering overhanging leads.
In figure~\ref{wires:fig2} $G(T)$ is shown for a 
{\it spinless} tight-binding wire with nearest-neighbor interaction $U'$ coupled to two 
semi-infinite leads via tunneling hoppings $T_{\rm L/R}$. 
For spinless models the low-energy LL physics can be described by $K$ and the charge
velocity $v_{\rm c}$. Several correlation functions are again characterized by power-law 
scaling. Compared to models with spin the analytic dependence of exponents on $K$ is 
modified (see e.g. \cite{Schoenhammer05}).   
Three setups with different overhanging 
parts are considered (see the left inset). As becomes clear each 
non-zero number of overhanging lattice sites $N_{\rm lead}^{\rm L/R}$ infers 
a new energy scale $\pi v_{\rm F}/N_{\rm lead}^{\rm L/R}$ and the simple power law (with 
exponent $\alpha_{\rm end}$; dashed line) is divided in subsections separated by extensive crossover 
regimes. If clearly developed the power laws in the subsections all have exponent $\alpha_{\rm end}$ (see the 
inset of figure~\ref{wires:fig2}). 
Overhanging parts of the wire lead to the same effect \cite{Waechter09}. There is no reason to believe that 
adding the spin degree of freedom will change this. The same holds for the two effects discussed next and 
we thus stick to the spinless lattice model for the rest of this section. 
  
One can conclude that in a generic experimental setup with overhanging parts the power law is 
much more difficult to observe than suggested by most theoretical studies
considering end-to-end coupling.

\subsection{A wire with two contacts and a single impurity}     

\begin{figure}[t]
  \centering
  \includegraphics[width=85mm,clip]{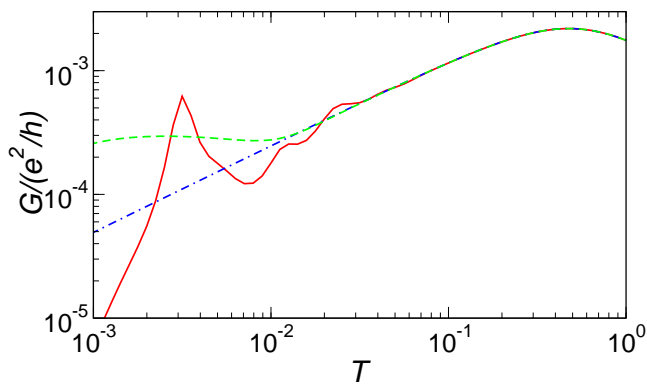}
  \caption{(Color online)  Linear conductance $G$ of an interacting tight-binding 
wire with tunnel couplings $T_{\rm L}=T_{\rm R}=0.1$, an additional hopping impurity (single bond with reduced hopping) of 
strength $t'=0.1$, and nearest-neighbor interaction $U'=1$ of length $N=10^4$ as function 
of the temperature $T$. The solid and dashed curve show data obtained for different positions of the bulk impurity. 
Dashed-dotted line: result obtained by phase averaging.}
  \label{wires:fig3}
\end{figure}

We now exemplify the very rich interference effects which occur when 
considering coherent linear transport through interacting wires with several inhomogeneities. 
A simple 
setup of this class is a wire with two end-to-end 
tunneling contacts to semi-infinite leads and an additional impurity 
in its bulk. As mentioned in the last subsection the concept of ``adding resistances'' cannot be applied in the
present context. Using averaging over the (two) {\it scattering phases} one can show that 
for sufficiently large $T$ (of the order $10^{-2}$ in figure~\ref{wires:fig3}) the total linear conductance is given by 
$G \sim \sqrt{G_1 G_2 G_3}$ with the conductances $G_i$ of the individual barriers. For the present 
setup we have $G_{1/3}(T) \sim T^{\alpha_{\rm end}}$ and $G_2(T) \sim  T^{2 \alpha_{\rm end}}$, leading to 
$G(T) \sim  T^{2 \alpha_{\rm end}}$~\footnote{Accidentally the exponent $2 \alpha_{\rm end}$ would also follow from the 
{\it unjustified} adding of resistances. }. As is shown in figure~\ref{wires:fig3} the 
dashed-dotted curve obtained by phase averaging nicely follows the bare data (solid and dashed line) 
down to $T_{\rm p} \approx {\mathcal O}(10^{-2})$ and a power law with the expected exponent
develops for $T_{\rm p} \lessapprox T \ll D$.  For $T < T_{\rm p}$ details of the {\it relative} 
energy-level structure of the two 
``dots'' defined by the three barriers matter  and might even lead to resonances (see the solid line). 
The ``dots" energy-level structure is set by the position of the bulk impurity and the solid and dashed 
line in figure~\ref{wires:fig3} display data obtained for different positions. 
The example clearly shows that interference due to multiple inhomogeneities can set a lower bound to 
power-law scaling of the conductance.        

\subsection{Transport at finite bias voltages}

\begin{figure}[t]
  \centering
  \includegraphics[width=85mm,clip]{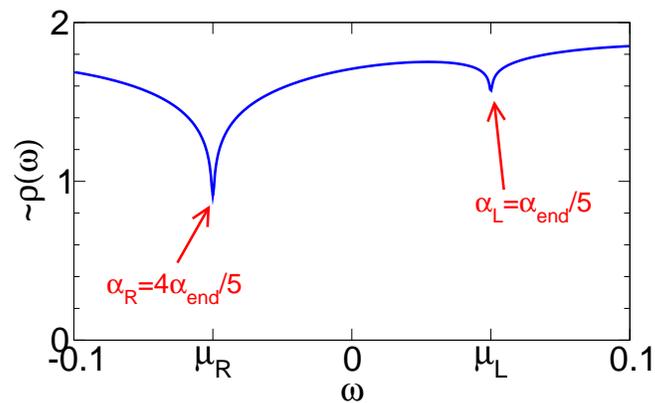}
  \caption{(Color online)   Solid line: Local spectral function (close to the left contact) in the 
steady-state of an interacting tight-binding wire of length $N=2 \cdot 10^4$ with tunnel couplings 
$T_{\rm L}=0.075$, $T_{\rm R}=0.15$, and nearest-neighbor interaction $U'=0.5$  as function 
of $\omega$. The voltage is $V = 0.1$}.
  \label{wires:fig4}
\end{figure}

As our final example we study transport at {\it finite bias voltages}. In particular we 
consider the steady state current through a clean wire end-to-end coupled to two leads (reservoirs) 
which are hold on different electrochemical potentials $\mu_{\rm L/R}=\pm V/2$. 
Applying a generalization of the fRG to non-equilibrium transport \cite{Jakobs07}  one 
can show that in the limit of high tunnel barriers the effective single-particle potential generated 
by the interplay of the contacts and the interaction is a {\it superposition} of two decaying oscillations 
with wave-lengths set by $\mu_{\rm L}$ and $\mu_{\rm R}$. The respective amplitudes are proportional to 
$T^2_{\rm L/R}/(T_{\rm L}^2 + T_{\rm R}^2) = \Gamma_{\rm L/R}/(\Gamma_{\rm L} + \Gamma_{\rm R})$, with 
$\Gamma_\alpha \sim T_\alpha^2$ defined as in the preceding sections.  
Accordingly, the non-equilibrium local spectral function 
close to either of the two contacts displays power-law suppressions at $\mu_{\rm L}$ {\it and} 
$\mu_{\rm R}$, $\rho(\omega) \sim |\omega-\mu_{\rm L/R}|^{\alpha_{\rm L/R}}$, with 
exponents proportional to the respective couplings $\alpha_{\rm L/R} = \Gamma_{\rm L/R} \alpha_{\rm end}/ 
(\Gamma_{\rm L} + \Gamma_{\rm R})$. In contrast to the linear response regime the exponents now dependent on the 
strength of the inhomogeneities. This behavior is shown in figure~\ref{wires:fig4}. The spectral weight 
remains finite even at $\mu_{\rm L/R}$ as the power-laws are cut off by the single-particle level spacing 
$\sim 1/N$ of the disconnected wire. A transport geometry in which the two power laws show up was proposed 
in \cite{Jakobs07}. 
 
\subsection{Outlook}

This short review on correlation effects in 1d quantum wires gives a flavor of the rich and partly surprising 
physics resulting from the interplay of the two-particle interaction and local (single-particle) 
inhomogeneities. An important ingredient missing in the current description using the (approximate) fRG 
for microscopic models are {\it inelastic scattering processes} resulting from the (Coulomb) interaction and leading 
to decoherence and dephasing at intermediate to large energies. We note that they are also only partly 
kept in the standard modeling which is based on the Tomonaga-Luttinger model. 
One can expect that inelastic processes set an upper energy scale $T_{\rm c}$ 
for power-law scaling smaller than the maximal scale given by the band width. 
They will also play a prominent role when further 
considering non-equilibrium transport. First attempts to include such inelastic two-particle 
scattering for quantum dots in and out of equilibrium  were mentioned in section \ref{sec:fluct_linear} and are discussed in \cite{Hedden04,Karrash08,Jakobs10}.

\section*{List of symbols}
\label{sec:list_of_symbols}

In this section we provide a list of the most relevant symbols used:

  \begin{longtable}{l|l}
    $\Delta\epsilon$ & level spacing 	\\
    $\Gamma$ & tunnel coupling \\
    $V_{\mathrm{g}}$ & gate voltage \\
    $V$ & bias voltage \\
    $dI/dV$ & differential conductance \\
    $t$ & time\\
    $T_{\mathrm{K}}$ & Kondo temperature \\
    $U$ & charging energy \\
    $\hbar=k_{\mathrm{B}}=e=1$	& units\\
    $T$ & temperature \\
    $|s\rangle$ & exact many-body level spectrum and states\\
    $H_D$ & dot Hamiltonian   \\
    $H_T$ & tunneling Hamiltonian  \\
    $T^{s s'}_{\alpha k\sigma}$ & tunnel amplitudes \\
    $\sigma$ & spin\\
    $\alpha=\mathrm{L,R}$ & reservoirs\\
    $H_{\alpha}$ & reservoir Hamiltonian   \\
    $n_{\alpha k \sigma}$ & electron number\\
    $k$ & label of the orbital state \\
    $\rho_\alpha$ & density of states \\
    $\mu_\alpha$ & electro-chemical potential\\
    $p(t)$ & dot density operator \\
    $L_{\mathrm{D}}$ & dot Liouville operator \\
    $\Sigma(t,t')$ & transport kernel \\
    $I_{\alpha}$ & tunnel current \\
    $\epsilon_\sigma$ & level energy \\
    $N$ & charge number\\
    $\omega$ & frequency\\
    $D_N$ & magnetic parameter \\
    $\lambda$ &  electron vibration coupling \\
    $\Omega$ & dwell time of electrons in the system \\
    $Q$ & pumped charge \\
    ${\mathcal{G}}_{(0)}$ &  (bare) Green function of the dot\\
    $\Sigma (i\omega)$ & self-energy \\
    $G$ & conductance\\
    $\delta_i$ & distance to resonances\\
    $z_i$ & poles of the reduced dot density matrix $\tilde{p}$ in Laplace space\\
    $h$ & magnetic field\\
    $J_{z/\perp}$ & exchange couplings \\
    $W^s$ & rates for the process $-s\rightarrow s$ \\
    $\chi$ & susceptibility \\
    $\Lambda$ & flow parameter \\
    $\Lambda_{\mathrm{c}}$ & cutoff \\
    $K$ & Luttinger liquid parameter\\
    $v_{\rm c/s}$  & velocities of the charge/spin-density excitations \\
    $k_{\rm F}$, $v_{\rm F}$ & Fermi momentum, Fermi velocity\\
    $U'$ & nearest-neighbor interaction\\
    $\alpha_{\rm bulk}$ & power-law exponent of the wire spectral function \\
    $\alpha_{\rm end}$ & power-law exponent of an electron tunneling into the end of a LL\\
  \end{longtable}

%Acknowledgments
%
\ack
We acknowledge all our collaborators in the various works presented in this review.
This work is supported by the DFG-FG 723, FG 912, SPP-1243, the NanoSci-ERA, the Ministry of Innovation NRW,
the Helmholtz Foundation and the FZ-J\"ulich (IFMIT).

\section*{References}
\bibliographystyle{unsrt}
\bibliography{paper}

\end{document}